\documentclass[12pt]{article}

\usepackage{textcomp}
\usepackage{chetnew}
\usepackage{amssymb,amsmath,amsfonts,mathrsfs,braket}
\usepackage[utf8]{inputenc}
\usepackage{hyperref,url}
\usepackage{tikz,tikz-cd}
\usepackage{xcolor}
\usepackage[boxruled]{algorithm2e}
\usetikzlibrary{arrows,decorations.markings,shapes.geometric,decorations.pathmorphing}
\tikzset{snake it/.style={decorate, decoration=snake}}

\usepackage{bm}
\usepackage[colorinlistoftodos,textsize=tiny]{todonotes}

\renewcommand{\d}[1]{\ensuremath{\operatorname{d}\!{#1}}}

\def\one{{\,\hbox{1\kern-.8mm l}}}

\allowdisplaybreaks

\def\makeatletter{\catcode`\@=11}% 11:letter
\makeatletter
\def\mathbox#1{\hbox{$\m@th#1$}}%
\def\math@ccstyles#1#2#3#4#5#6#7{{\leavevmode
      \setbox0\mathbox{#6#7}%
      \setbox2\mathbox{#4#5}%
      \dimen@ #3%
      \baselineskip\z@\lineskiplimit#1\lineskip\z@
      \vbox{\ialign{##\crcr
             \hfil \kern #2\box2 \hfil\crcr
             \noalign{\kern\dimen@}%
             \hfil\box0\hfil\crcr}}}}
\def\mathaccstyles{\math@ccstyles\maxdimen}
\def\maththroughstyles{\math@ccstyles{-\maxdimen}}
\def\unity%
 {\maththroughstyles{.45\ht0}\z@\displaystyle {\mathchar"006C}\displaystyle 1}

\def\II{{\cal I}}

\def\MM{{\cal M}}
\def\NN{{\cal N}}
\def\OO{{\cal O}}

\def\II{{\mathbb I}}

\def\dmax{{\Delta_{\rm max}}}

\def\E{{\mathbf{E}}}

\def\d{{\partial}}

\def\C{{\mathfrak C}}
\def\A{{\mathbb A}}

\def\beq{\begin{equation}}
\def\eeq{\end{equation}}
\newcommand{\bea}{\begin{eqnarray}}
\newcommand{\eea}{\end{eqnarray}}
\def\bal{\begin{align}}
\def\eal{\end{align}}

\preprint{QMUL-PH-21-34\\ ITCP-IPP 2021/2\\ CCTP-2021-4}

\title{\vspace{-0.cm}Conformal Bootstrap with Reinforcement Learning
}

\author{Gergely K\'antor\;$^{a,\spadesuit}$, Vasilis Niarchos\;$^{b,\diamondsuit}$ and Constantinos Papageorgakis\;$^{a,\clubsuit}$}

\affiliation{$^a$Centre for Theoretical Physics, Department of Physics and Astronomy\\ Queen Mary University of London, London E1 4NS, UK \vspace{0.3cm} $ $ \\

$^b$CCTP and ITCP, Department of Physics,\\
University of Crete, 71003 Heraklion, Greece
\vspace{0.3cm} $ $\\

\vspace{0.3cm}
{\tt \small$^\spadesuit$g.kantor@qmul.ac.uk, $^\diamondsuit$niarchos@physics.uoc.gr, $^\clubsuit$c.papageorgakis@qmul.ac.uk}}

\abstract{We introduce the use of reinforcement-learning (RL) techniques to the conformal-bootstrap programme. We demonstrate that suitable soft Actor-Critic RL algorithms can perform efficient, relatively cheap high-dimensional searches in the space of scaling dimensions and OPE-squared coefficients that produce sensible results for tens of CFT data from a single crossing equation. In this paper we test this approach in well-known 2D CFTs, with particular focus on the Ising and tri-critical Ising models and the free compactified boson CFT. We present results of as high as 36-dimensional searches, whose sole input is the expected number of operators per spin in a truncation of the conformal-block decomposition of the crossing equations. Our study of 2D CFTs uses only the global $so(2,2)$ part of the conformal algebra, and our methods are equally applicable to higher-dimensional CFTs. When combined with other, already available, numerical and analytical methods, we expect our approach to yield an exciting new window into the non-perturbative structure of arbitrary (unitary or non-unitary) CFTs.}

\date{}

\begin{document}

\maketitle

\hypersetup{pageanchor=true}

\setcounter{tocdepth}{2}

\toc

%%%%%%%%%%%%%%%%%%%%%%%%%%%%%
\section{Introduction}
\label{intro}

The non-perturbative formulation of a generic Quantum Field Theory (QFT) and the analytic, or numerical, solution of its dynamics remains an extremely challenging conceptual and computational problem with important theoretical and experimental implications.

The problem becomes more tractable in Conformal Field Theories (CFTs): a special class of QFTs that describe typically the short and large-distance behaviours of generic QFTs. Most notably, in a unitary, relativistic CFT in $D$ spacetime dimensions, the local structure of the theory is characterised by a set of discrete data: the scaling dimensions $\Delta_i$ of local conformal primary operators $\OO_i$ and their Operator Product Expansion (OPE) coefficients $C_{ij}^k$. Once these data are known, the generic correlation function of any local operator in the theory can be determined.

Unitarity implies certain well-known constraints on these data. For example, a conformal primary operator with scaling dimension $\Delta$ and spin $s$ must satisfy the inequalities
\bea
\label{introaa1}
&&\Delta \geq \frac{D-2}{2}~, ~~{\rm for}~ s=0
\\
\label{introaa2}
&&\Delta \geq s + D-2~, ~~ {\rm for} ~ s>0
~.
\eea
The equality $\Delta = s+D-2$ occurs only for conserved currents.

More elaborate, and powerful, constraints on the CFT data arise from crossing symmetry: the property that a correlation function is the same irrespective of the channel used in its OPE decomposition. These constraints (consistency conditions) form the basis of the {\it conformal bootstrap approach}. Since the 1970s (see e.g.~\cite{Ferrara:1973yt,Polyakov:1974gs}) it was hoped that by solving the conformal bootstrap equations, one would be able to solve CFTs non-perturbatively, without the need for a Lagrangian formulation. For many years the complexity of the conformal bootstrap equations, and the fact that they admit an infinite set of solutions for an infinite set of unknowns, did not allow the programme to evolve beyond a limited set of cases in 2D conformal field theory.

\subsection{Brief Background on the Modern Conformal Bootstrap}

Significant progress was instigated in 2008 by the seminal paper \cite{Rattazzi:2008pe}, which shifted the focus away from the search of exact solutions of the conformal bootstrap equations and towards the following approach: {\it Make an assumption about the spectrum of the CFT and ask if the bootstrap equations can be satisfied; if the equations cannot be satisfied, then this assumption can be successfully eliminated.} With suitable truncations on the infinite-dimensional CFT spectrum, this programme can be implemented numerically, and powerful linear and semidefinite programming methods\footnote{A commonly used package is the Semidefinite Program Solver (SDPB) \cite{Simmons-Duffin:2015qma,Landry:2019qug}.} have been employed in recent years to obtain many significant results in this direction. It is impossible to list here all the results and different applications of this approach. For a concise review, and orientation to the relevant literature, we refer the reader to \cite{Simmons-Duffin:2016gjk,Poland:2018epd,Chester:2019wfx}.

The assumptions that drive this approach are selected blindly; in the words of \cite{Reehorst:2021ykw}, the bootstrap computations in this context are performed in an ``oracle mode''. Nevertheless, suitable assumptions not only carve out significant parts of the space of potential CFTs, but one interestingly finds in many cases that known theories lie at cusps of the boundary of allowed possibilities. Even more efficiently, sometimes one discovers that the allowed region is an isolated ``island''. When this happens, the oracle-mode can be used to compute remarkably well scaling dimensions and OPE coefficients. A beautiful application of this method is encountered in the 3D Ising model \cite{ElShowk:2012ht,Kos:2014bka}. Theories at the boundary of the allowed and disallowed regions are obviously special from this perspective and have been the primary target of standard applications of the conformal bootstrap. Efficient computational methods, like the Extremal Functional Method \cite{El-Showk:2012vjm,El-Showk:2016mxr}, can be used to enhance the arsenal of the conformal bootstrap in this context.

Nevertheless, some obvious shortcomings of this approach include:
\begin{itemize}

\item[$(a)$] For theories inside the allowed region one cannot, in general, tell how far they are located away from the boundary.

\item[$(b)$] With generic assumptions in oracle mode it is hard to identify and solve specific pre-selected CFTs, such as one's favourite gauge (conformal field) theory, that may not lie on the boundary of allowed and disallowed regions of the search.

\item[$(c)$] Higher-dimensional searches that would facilitate the study of more general classes of CFTs are computationally expensive and difficult to implement with the existing techniques. Typically, with current standard techniques one is restricted to searches of a couple of parameters.

\end{itemize}

To address some of these problems Ref.\ \cite{Reehorst:2021ykw} recently introduced the {\it Navigator-function} method, which replaces the binary information of the oracle mode with continuous information from an optimised continuous, differentiable function, called the Navigator function. The Navigator function is positive in the disallowed region, negative in the allowed region, zero at the boundary and, in principle, it is defined globally on the space of parameters. By minimising the Navigator function one can flow from a disallowed region to an allowed region and thus map out islands in parameter space, e.g.~by finding one feasible point inside the island or by finding an island's extremal points. The algorithms of \cite{Reehorst:2021ykw} employ the same well-developed semi-definite programming tools of SDPB that were previously used to determine OPE coefficients as a maximisation problem. Notable precursors of the Navigator-function method are the optimisation methods proposed in \cite{Poland:2010wg,Afkhami-Jeddi:2019zci}.

Another notable approach to the conformal bootstrap, with the potential to address the above issues, was proposed earlier on by Gliozzi in \cite{Gliozzi:2013ysa}; see also \cite{Gliozzi:2014jsa,Gliozzi:2015qsa,Gliozzi:2016cmg,Esterlis:2016psv,Hikami:2017hwv,Hikami:2017sbg,Li:2017ukc,Leclair:2018trn} for further work in this direction. In \cite{Gliozzi:2013ysa} the conformal-block expansion of the crossing equations was arbitrarily truncated and Taylor-expanded in cross-ratio space. A specific assumption was made about the spectrum of operators that enter the truncated conformal-block expansions. Viewing the resulting crossing equations as an over-constrained system of linear equations for the unknown OPE-squared coefficients, and demanding the existence of non-trivial solutions, yields conditions on the allowed scaling dimensions, which are phrased as vanishing determinants. This method can be used, in principle, to study a wider class of CFTs, including non-unitary CFTs, which are beyond the reach of the above-mentioned SDPB approaches. It requires, however, that the CFT is ``truncable'', which is not an a priori obvious property of a given CFT (see \cite{Gliozzi:2014jsa} for an example that is not truncable). In \cite{Li:2017ukc}, the Gliozzi approach was reformulated as a minimisation problem, which improves important aspects of the method. The approach to the conformal bootstrap that we introduce in this paper is similar in spirit to the reformulation of \cite{Li:2017ukc}.

Both of the above approaches, and the one we introduce below, are phrased as optimisation problems. A distinctive feature of what we do is that instead of minimising directly the quantity of interest, we optimise a Neural Network (NN) that predicts a probability distribution, which is then sampled to make the actual predictions. This approach has several advantages. In direct optimisation function, one needs to compute partial derivatives, which can become expensive in high-dimensional searches.\footnote{In \cite{Reehorst:2021ykw} this problem is avoided with a general SDP gradient formula and the efficient use of a quasi-Newton method.}
In contrast, we use fixed optimisation algorithms for the NNs, independent of the details and complexity of the specific problem. Moreover, when one optimises  the function of interest directly, one has to first pick a point in state-space to initialise the process, and then the derivatives guide the search towards the closest minimum. In order to flow to the minimum, one has to pick a small enough learning rate, but that inevitably restricts the flow to the closest minimum, even if it is not the global one. Our approach is efficient at trying to find the global minimum, because the learning rate varies and it probes minima at varying distances from the original starting point. The price we have to pay for these advantages is that our computations become less ``exact'', i.e.~less direct and more statistical.

\subsection{A Novel Study of Truncations Based on Artificial Intelligence}
\label{aiintro}

In the present work we study truncated crossing equations as an optimisation problem and develop methods to find approximate numerical solutions taking advantage of recent developments in Machine Learning (ML) and the wider availability of associated techniques. Similar to \cite{Gliozzi:2013ysa,Li:2017ukc}, our approach is more akin to the original philosophy of the 1970s, which aimed at a {\it direct solution} of the conformal bootstrap equations. We will explain momentarily how we set up and implement a multi-dimensional search of approximate solutions and how this search benefits from artificial intelligence-techniques.

\subsubsection{Introductory Comments on ML Terminology}

Designing architectures and algorithms which one day could surpass human performance has been a long-running goal in the field of ML. Although a significant part of the theoretical (statistical and probabilistic) groundwork had been laid down for more than half a century, ML has only recently started to truly flourish. Decades ago algorithms which beat professional chess players had already been designed, but these approaches involved codes that were rigid and non-dynamic, meaning that once written their knowledge would be capped. In contrast, all of the modern developments in having machines learn how to solve problems include dynamic programming and a statistical approach to learning. The latter has only become practically feasible of late with the rapid development of and easier access to powerful central processing units (CPUs) and graphics processing units (GPUs).

The three best-known categories of ML algorithms are: {\it supervised}, {\it unsupervised} and {\it reinforcement learning}. In supervised learning some of the data are tagged and contain both the input and desired output. The algorithm trains on the tagged data and learns how to produce a sensible output from any input. Typical applications of supervised learning are classification and regression problems. In unsupervised learning there are no externally provided tagged data for training; the algorithm recognises on its own structure in a given set of data. In Reinforcement Learning (RL) \cite{sutton2018reinforcement}---or Deep Reinforcement Learning (DRL), that employs Deep Neural Networks (DNNs) in the learning steps of the ``agent''---one knows the goals but does not know how to achieve them. The algorithm interacts with a dynamic environment and receives feedback based on its performance that guides it towards the desired result. 

In recent years, ML has had a rising number of applications in High Energy Physics.\footnote{See \cite{MLHighEnergy,MLLQCD} for a compendium of reviews ranging from the more experimental to the more computational aspects, and \cite{Ruehle:2020jrk} for a summary of applications to String Theory. RL implementations have appeared in the context of String Theory even more recently in \cite{Halverson:2019tkf,Harvey:2021oue,Krippendorf:2021uxu,Constantin:2021for}. See also \cite{Tanaka} for a nice introduction to deep learning from a physics-motivated viewpoint.} In this paper, we will initiate a study of the conformal-bootstrap programme using RL techniques. This is the first study of conformal field theory of this kind.\footnote{An alternative ML approach towards certain aspects of CFT, using supervised learning, appeared in \cite{Chen:2020dxg}. The methodology, focus and scope of \cite{Chen:2020dxg} are very different from the one that we introduce below.}

\subsubsection{RL Setup in the Conformal Bootstrap}
\label{ingredients}

Ultimately, a successful RL algorithm should be able to {\it identify} a proper CFT, by converging to a configuration of CFT data that satisfy the crossing equations within a prescribed accuracy. It should similarly be able to {\it exclude} improper CFTs by failing to converge to a configuration that satisfies the crossing equations within the prescribed accuracy.

The basic scenario of our approach includes the following ingredients:
\begin{itemize}

\item Consider a specific four-point function with operators that have fixed symmetry properties, scaling dimensions and spins. If the scaling dimensions of the external operators are unknown, one can include them, as unknown variables, into the search.

\item The crossing equations are truncated with a specific assumption about the number of operators per spin that appear in each channel. We call this assumption the {\it spin-partition} of the truncated conformal-block expansion. For example, if the truncation of the conformal block expansion in a given channel is assumed to include only operators of integer spin, and we truncate at maximum spin 3, then the spin-partition specifies the number of operators at spin 0, 1, 2 and 3. The spin-partition, which is an input to the RL algorithm, specifies the dimensionality of the vector of unknown scaling dimensions and OPE-squared coefficients $( \vec \Delta, \vec \C)$, that we aim to determine.

\item We assume that the conformal blocks are known analytically, or numerically,  \cite{Simmons-Duffin:2016gjk,Poland:2018epd,Chester:2019wfx}. The crossing equations, which are functions of the cross-ratios (see Sec.~\ref{generalities} for details), are reduced to a set of algebraic equations for the unknown scaling dimensions and OPE-squared coefficients $( \vec \Delta, \vec \C)$. The reduction can be achieved by Taylor expanding the conformal blocks around a particular point (as in standard applications of the numerical conformal bootstrap), or by evaluating the conformal blocks on a set of different points in cross-ratio space. We will implement the latter approach in this paper. Naturally, the number of algebraic crossing equations obtained in this manner should be larger than the number of unknowns. In compact vector form, the reduced algebraic crossing equations are
\beq
\label{RLbasicaa}
\vec \E (\vec \Delta, \vec \C) = 0
~.
\eeq
Since we truncate the crossing equations, it is not guaranteed (or expected) that Eqs.~\eqref{RLbasicaa} have an exact solution. Our aim is to find approximate solutions to \eqref{RLbasicaa} that minimize $\vec \E$. Approximate solutions are expected to flow towards exact solutions of the exact crossing equations as one adds more and more operators to the truncation.

\item One can specify the width of the search either individually for each unknown scaling dimension and OPE-squared coefficient, or collectively. For example, one can set a common upper cutoff, $\dmax$, on the unknown scaling dimensions. Clearly, because of the unitarity constraints, \eqref{introaa1}-\eqref{introaa2}, if the maximum spin in the spin-partition is $s_{\rm max}$, then $\dmax \geq s_{\rm max}+D-2$.

\item With these specifications in mind, we set up a soft Actor-Critic RL algorithm, \cite{DBLP:journals/corr/abs-1801-01290}, that performs a multi-dimensional search on the vector space of the unknown scaling dimensions and OPE-squared coefficients $( \vec \Delta, \vec \C)$ and returns configurations that minimise the norm of the crossing-equation vector $\vec \E$. The operation and key components of the RL algorithm will be discussed in Sec.~\ref{ml}.

\end{itemize}

\subsection{Overview and Discussion of Results}
\label{sumresults}

Our main goal in this paper is to show that suitable RL algorithms can be applied to the conformal-bootstrap programme to efficiently perform multi-dimensional searches, and (when appropriately guided) to detect and solve arbitrary CFTs. We aim primarily at a proof-of-concept demonstration of the approach with less emphasis on maximising the accuracy of the results, which we will consider in future work. In that vein, we want to test RL algorithms against results that can be obtained independently using analytic methods.

We choose to analyse 2D CFTs, as in this case it is straightforward to write exact conformal blocks for operators of arbitrary spin. Throughout our computations, we will only use the global $so(2,2)$ part of the 2D conformal algebra, without making any reference to the Virasoro algebra, which is a special feature of two dimensions. Consequently, every tool that we set up in this paper is directly generalisable and applicable to higher-dimensional CFTs, which will be treated elsewhere. For concreteness, we will focus separately on the two leading unitary minimal models (the Ising and tri-critical Ising model) and the free boson CFT on a circle.

\subsubsection{Key Results}

We highlight the following results:

\begin{itemize}

\item In all the cases we analysed, the algorithm was able to detect the CFT whose spin-partition we used as input. This is extremely promising. It suggests that Reinforcement Learning has a great potential as a tool in conformal-bootstrap studies of generic pre-selected CFTs. Our approach is not limited to special theories, e.g.\ CFTs on cusps of parameter spaces, or CFTs with enhanced symmetries.

\item Even with a relatively small upper cutoff on the scaling dimensions our algorithm produces sensible numerical results that satisfy the truncated crossing equations at good accuracy. The details depend on the theory and the four-point function that we are analysing. For instance, for simple CFTs like the 2D Ising model, a run with only 5 quasi-primary operators yields scaling dimensions and OPE-squared coefficients comparable with their analytic values within the order of 1\%. In the free compactified boson CFT we obtain sensible results even with 4 quasi-primary operators and cutoff $\dmax=2$. As one might expect, the results of our RL algorithm are generically more accurate for lower scaling dimensions, and less accurate for quasi-primaries close to the cutoff when compared with the analytic answers.

\item We can probe the dependence of CFTs on closely spaced discrete parameters, or continuous parameters like exactly marginal couplings. We present examples of such a study in the context of the 2D free boson on a circle. In that case, the continuous parameter is the radius of the circle. Being applicable in such scenarios, our method could readily be combined with analytic results in convenient parameter regimes (e.g.\ at weak-coupling points) to solve the theory at generic points by adiabatically changing the parameters.

\item We can perform efficient high-dimensional searches; our current algorithm can do direct searches with tens of operators. In the context of the 2D compactified boson CFT, we present results of a run with 36 parameters. We can, in principle, go to even higher spins and scaling dimensions with multiple sequential runs that start with a smaller number of operators and gradually introduce more.

\end{itemize}

\subsubsection{Numerical Uncertainties}

An important aspect of our approach, which is not addressed in detail in the preliminary investigations of this paper, has to do with the systematic treatment of errors. As emphasised at the beginning of this subsection, the main goal of the present work is to establish that our algorithm detects the intended CFT and produces sensible numbers. We achieve this goal by comparing said numbers with the available exact analytic results. A preliminary discussion of errors and uncertainties, and how they can be incorporated systematically in the future, is relegated to the concluding Sec.~\ref{outlook}.  In the rest of this subsection, we flesh out an important aspect of our approximations that affects the implementation of our approach.

As already noted, the truncated crossing equations that we are trying to solve do not admit, in general, any exact solutions. Therefore, our main task is to find configurations that minimise the violation of the truncated equations. What is the minimal violation of the truncated equations that we should be aiming for? This is not a priori known and the answer can depend strongly on the specifics of the CFT, the four-point function that we are considering, the type of truncation that we are implementing on the spectrum and the way we reduce the crossing equations as functions in cross-ratio space to a number of algebraic equations. The answer to this question has obvious practical implications. Most notably, it determines when a run should be terminated and affects the decision of whether a given output should be accepted as a solution to an actual CFT, or whether it should be rejected as a false minimum.

In Sec.~\ref{defspinpartitions} we define a measure of relative accuracy $\A$ (see Eqs.\ \eqref{defsspinsac}, \eqref{defsspinsad}) that quantifies a $\%$ violation of the truncated crossing equations. $\A$ has a minimum value $\A_{\min}$ for searches in a compact subspace of parameter space. It is expected that $\A_{\min} \to 0$ as we incorporate more and more operators, but it is not obvious, in general, how to determine $\A_{\min}$ as a function of all the factors that were listed in the previous paragraph. If there is a regime, where the analytic solution is known, $\A_{\min}$ can be estimated with a direct RL-algorithm run in the vicinity of the known solution. This estimate can then be used as a guide in other regimes of parameters where the analytic solution is not known.

We have empirically found that in all computations performed for this paper a solution has been properly identified for values of $\A$ below $0.1\%$ irrespective of the spin truncation. Once $\A$ is below this empirical threshold {\it and} $\A$ stops improving {\it and} the agent has visibly converged to a configuration, we terminate the run and record the result. We have implemented this triple selection rule in all the runs that are reported in this paper.

To obtain further evidence for the acceptance, or rejection, of a configuration one can study the dependence of the best $\A$ obtained by the algorithm as more and more operators are included. Once a configuration has been accepted as a valid approximation to the exact problem, one can define individual uncertainties for each CFT datum that is being computed. We present preliminary results of statistical errors in specific examples in Sec.~\ref{boson}. We discuss general uncertainties and their sources further in the concluding Sec.~\ref{outlook}.

\subsection{Outline}
\label{plan}

The rest of this paper is organised as follows. In Sec.~\ref{mltheory} we present a brief review of useful basic CFT properties and set up our notation. We introduce the truncation scheme that we use, the associated spin partitions and a measure of accuracy that plays a key role in the numerical computations of the main text. In Sec.~\ref{ml} we summarise the main features of continuous action space Reinforcement Learning. We describe the key components of the soft Actor-Critic algorithm and outline three practical modes of implementation. Secs~\ref{minimal} and \ref{boson} are the central sections of the paper. In Sec.~\ref{minimal} we present an RL study of four-point functions of the spin and energy-density operators in the 2D Ising and tri-critical Ising models. In Sec.~\ref{boson} we study four-point functions of primary operators in the momentum/winding sector of the compactified boson CFT and four-point functions of the conserved $U(1)$ current. We discuss the dependence of the results on the scaling dimension cutoff $\dmax$ and the exactly marginal coupling of the theory. We conclude in Sec.~\ref{outlook} with a brief synopsis of the main results and an outlook on future directions. 

A shorter version of this paper, summarising the key approaches and results, can be found in \cite{MLshort}.

%%%%%%%%%%%%%%%%%%%%%%%%%%%%

\section{CFT Prerequisites and Notation}
\label{mltheory}

In what follows we assume some familiarity with the basic concepts of conformal field theory. For a review of conformal field theory we refer the reader to the standard textbook \cite{DiFrancesco:1997nk} and the recent overviews in \cite{Simmons-Duffin:2016gjk,Poland:2018epd,Chester:2019wfx}, which summarise the more modern perspective on CFTs above two dimensions. Sec.~\ref{generalities} provides a general overview of useful properties for CFTs in any spacetime dimension. In Secs~\ref{2dintro} and \ref{defspinpartitions} we specialise the discussion to 2D CFTs, which will be the main focus of the computations in this paper.

\subsection{Generalities}
\label{generalities}

The $so(D,2)$ conformal algebra of a CFT in $D$ spacetime dimensions organises the spectrum of local operators/states of the theory in corresponding representations. A primary operator $\OO_i$ has scaling dimension $\Delta_i$ and spin (under the $SO(D)$ Lorentz group) $s_i$. Notice that the case $D=2$ is special, since the $so(2,2)$ part of the conformal algebra extends to the infinite-dimensional Virasoro algebra. It is, therefore, customary in 2D CFTs to refer to the operators that are highest-weights in Virasoro representations as primaries, while operators that are highest-weights in representations of the global part $so(2,2)$ are called quasi-primary. Since we will be using only the $so(2,2)$ structure of 2D CFTs, the reader should anticipate a clear distinction between primary and quasi-primary operators in the context of our applications.

A central object in the analysis of CFTs is the Operator Product Expansion (OPE), which allows one to recast the product of two conformal primaries $\OO_i$, $\OO_j$ as a sum over single conformal primaries and their descendants
\beq
\label{mltheoryaa}
\OO_i(x_1) \OO_j(x_2) = \sum_k C_{ij}^k \hat f_{ij}^k \left(x_1,x_2,\d_{x_2} \right) \OO_k(x_2)
~.
\eeq
The OPE coefficients $C_{ij}^k$ are c-numbers that are closely connected to the three-point function coefficients $C_{ijk}$ of the conformal primaries $\OO_i, \OO_j, \OO_k$. For example, the two- and three-point functions of three conformal primary scalar operators are given by the expressions
\beq
\label{mltheoryab}
\langle \OO_i (x_1) \OO_j(x_2) \rangle = \frac{g_{ij}}{|x_{12}|^{2\Delta}}~, ~~ {\rm for}~~ \Delta_i = \Delta_j \equiv \Delta
~,
\eeq
\beq
\label{mltheoryac}
\langle \OO_i (x_1) \OO_j(x_2) \OO_k(x_3) \rangle = \frac{C_{ijk}}{|x_{12}|^{\Delta_{ij,k}} |x_{23}|^{\Delta_{jk,i}} |x_{13}|^{\Delta_{ik,j}} }
~,
\eeq
with $\Delta_{ij,k}\equiv \Delta_i + \Delta_j - \Delta_k$ and $x_{ij}=x_i-x_j$. In this case, $C_{ijk} = \sum_k C_{ij}^m g_{mk}$. The conformal symmetry forces the two-point functions in \eqref{mltheoryab} to vanish if $\Delta_i \neq \Delta_j$ and fixes the spacetime dependence of both the two- and three-point functions. For spinning operators the expressions in \eqref{mltheoryab}, \eqref{mltheoryac} generalise to include the tensor structure of the spins. The quantity $\hat f^k_{ij}$ in the sum \eqref{mltheoryaa} is a differential operator that incorporates the contributions of all the conformal descendants in the conformal multiplet of $\OO_k$. Its form is fixed by conformal symmetry.

The OPE \eqref{mltheoryaa} can be used to reduce a generic $n$-point function to a sum of products of three-point functions. Hence, the full dynamical content of local correlation functions in a CFT can be captured by the knowledge of two- and three-point functions. Equivalently, the solution of the local structure of a CFT entails the computation of the full spectrum of scaling dimensions $\Delta_i$ at each spin $s_i$ and of the corresponding OPE coefficients $C_{ij}^k$.\footnote{Another special feature of CFTs is the operator/state correspondence. We will frequently use it to interchange language between states and operators.}

Four-point functions $\langle \OO_{i_1} (x_1) \OO_{i_2} (x_2) \OO_{i_3} (x_3) \OO_{i_4}(x_4) \rangle$ provide a powerful demonstration of this reduction. Unlike \eqref{mltheoryab}, \eqref{mltheoryac}, conformal symmetry does not completely fix the spacetime dependence of four-point functions. Solely from the viewpoint of conformal symmetries we can write
\beq
\label{mltheoryae}
\langle \OO_{i_1} (x_1) \OO_{i_2} (x_2) \OO_{i_3} (x_3) \OO_{i_4}(x_4) = {\boldsymbol K}(\Delta_i,x_i) \, g(u,v)
~,
\eeq
where the factor ${\boldsymbol K}(\Delta_i, x_i)$ has a fixed form (that will be written explicitly in two dimensions below), and $g(u,v)$ is a---typically complicated---theory-specific function of the cross-ratios
\beq
\label{mltheoryad}
u = \frac{x_{12}^2 x_{34}^2}{x_{13}^2 x_{24}^2} ~,~~
v =  \frac{x_{14}^2 x_{23}^2}{x_{13}^2 x_{24}^2}\;,
\eeq
which are invariant under conformal transformations. The OPE expansion \eqref{mltheoryaa} of the products $\OO_{i_1}\OO_{i_2}$ and $\OO_{i_3} \OO_{i_4}$ allows us to recast \eqref{mltheoryae} as
\beq
\label{mltheoryaf}
\langle \OO_{i_1} (x_1) \OO_{i_2} (x_2) \OO_{i_3} (x_3) \OO_{i_4}(x_4) = {\boldsymbol K}(\Delta_i,x_i) \, \sum_{k_1,k_2} C_{i_1 i_2}^{k_1} g_{k_1k_2} C_{i_3 i_4}^{k_2} g_{\OO_k}^{(i_1i_2i_3i_4)}(u,v)
~,
\eeq
where $g_{\OO_k}^{(i_1i_2i_3i_4)}(u,v)$ is the conformal block that captures the contribution of intermediate operators $\OO_{k_1}$, $\OO_{k_2}$ with equal scaling dimension $\Delta_k$. The conformal blocks are theory-independent and, as already mentioned earlier, in many cases are either known analytically in closed form, or can be determined using convenient relations. Specific expressions for two-dimensional conformal blocks will be given momentarily.

It is customary (in the context of the so-called conformal frame) to re-express the cross-ratios in terms of two variables $z, \bar z$ as
\beq
\label{mltheoryag}
u = z \bar z~, ~~ v = (1-z)(1-\bar z)
~.
\eeq
In Euclidean CFT $z$ and $\bar z$ are complex conjugate.

It is also customary to work in a basis of conformal primaries that diagonalises the two-point functions \eqref{mltheoryab}. This is a convenient choice in general, but it can be subtle in conformal manifolds for degenerate protected operators because of operator-mixing effects. In what follows we denote the OPE-squared sum at fixed scaling dimension $\Delta_k$ as
\beq
\label{mltheoryai}
\C_{i_1 i_2 i_3 i_4}^k \equiv \sum_{k_1, k_2 \, | \, \Delta_{k_1}=\Delta_{k_2} = \Delta_k} C_{i_1 i_2}^{k_1} g_{k_1k_2} C_{i_3 i_4}^{k_2}
~.
\eeq
In the absence of degeneracies in the spectrum of operators that run in this sum, the sum \eqref{mltheoryai} comprises a single term. This is not, however, the only possibility and in some of the applications of the main text we will encounter cases where degeneracies do exist. Our algorithm tries to determine the full coefficients $\C_{i_1 i_2 i_3 i_4}^k$, hence if there are degeneracies it will not be able to resolve them to determine the individual contributions that make up the sum in \eqref{mltheoryai}.

Obviously, the OPE expansion in \eqref{mltheoryaf} is not unique. Instead of using the OPEs $\OO_{i_1}\OO_{i_2}$ and $\OO_{i_3} \OO_{i_4}$ one can use the OPEs $\OO_{i_3}\OO_{i_2}$ and $\OO_{i_1} \OO_{i_4}$ to obtain a different looking, but equivalent, expansion of the four-point function. These two approaches yield respectively the so-called $s$- and $t$-channel expansions of the four-point function.\footnote{It is also possible to consider the $(13)-(24)$ OPEs that yield the $u$-channel expansion. We will not consider the $u$-channel expansion in this paper. We note that the $s$, $t$ and $u$ channel expansions do not converge simultaneously at all cross-ratio values. For further comments we refer the reader to the review \cite{Poland:2018epd}.} To distinguish the OPE-squared coefficients in each channel, we will denote the $s$-channel coefficients as $_s \C_{i_1i_2i_3i_4}^k$ and the $t$-channel coefficients as $_t \C_{i_1i_2i_3i_4}^k$. The $t$-channel can be obtained from the $s$-channel by exchanging the insertions $1\leftrightarrow 3$ and equivalently the cross-ratios $u\leftrightarrow v$, or $z\leftrightarrow 1-z$ and $\bar z\leftrightarrow 1-\bar z$. The equality of the two expansions leads to the crossing symmetry constraints
\beq
\label{mltheoryaj}
{\sum_k}\, _s \C_{i_1i_2i_3i_4}^k g_{\Delta_k}^{(i_1i_2i_3i_4)}(u,v) - {\sum_{k'}}\, _t \C_{i_1i_2i_3i_4}^{k'}\, h(\Delta_i; u,v) g_{\Delta_k}^{(i_3i_2i_1 i_4)}(v,u) =0
~,
\eeq
where the factor $h(\Delta_i; u,v)$ accounts for the contribution of the prefactor $\boldsymbol K$.

In general, the operators that appear in the $s$-channel $k$-sum are different from the operators that appear in the $t$-channel $k'$-sum. Moreover, note that the crossing equations \eqref{mltheoryaj} have to be satisfied as functions of $u,v$ at any values of $u,v$. This imposes stringent constraints on the CFT data of scaling dimensions and OPE coefficients. We will set up an RL algorithm that solves these equations---yielding the CFT data---using an assumption about the rough structure of the spin-dependence of the spectrum of operators that appear in the OPE of each channel.

\subsection{Crossing Equations in 2D CFTs}
\label{2dintro}

It will be useful for our purposes to spell out the above results in the more specific case of two-dimensional CFTs.

The analysis of the crossing equations \eqref{mltheoryaj} requires explicit knowledge of the conformal blocks $g_{\Delta_k}^{(i_1i_2i_3i_4)}(u,v)$. Over the years significant progress in the computation of conformal blocks (see \cite{Poland:2018epd} for a guide to the literature) has provided important input in the development of the conformal-bootstrap programme. In even-dimensional CFTs the conformal blocks in four-point functions of scalar operators are known analytically in closed form. In two-dimensional CFTs, in particular, they are also known analytically for any four-point function of spinless or spinning conformal primary operator \cite{Osborn:2012vt}. The latter is one of the basic reasons why we will focus on 2D CFTs. We stress again that the aforementioned conformal blocks in two dimensions are conformal blocks for the global $so(2,2)$ part of the Virasoro algebra. In this paper we will not be using Virasoro conformal blocks.\footnote{In two dimensions it would have been more efficient, in general, to work with the full Virasoro blocks. However, this would be problematic for us for two reasons. First, the general Virasoro conformal blocks are not known in closed analytic form (see, however,  \cite{Perlmutter:2015iya} for useful expansions of these quantities). Second---and more important---this would limit the direct applicability of our approach to the special features of two-dimensional CFTs.}

Concretely, consider four quasi-primary operators in a (Euclidean) 2D CFT denoted as $\OO_i$ $(i=1,2,3,4)$ with left- and right-moving conformal weights $(h_i, \bar h_i)$. The corresponding scaling dimensions and spins of these operators are $\Delta_i = h_i+\bar h_i$ and $s_i = h_i - \bar h_i$. We insert the operators at four distinct spacetime points denoted in complex coordinates as $(z_i,\bar z_i)$. The $s$-channel conformal-block expansion of the four-point function of these operators is
\bea
\label{mltheoryba}
&&\langle \OO_1 (z_1,\bar z_1) \OO_2 (z_2,\bar z_2) \OO_3 (z_3,\bar z_3) \OO_4(z_4,\bar z_4) \rangle
=  \frac{1}{z_{12}^{h_1+h_2} z_{34}^{h_3+h_4}} \frac{1}{\bar z_{12}^{\bar h_1+\bar h_2} \bar z_{34}^{\bar h_3+\bar h_4}}
\nonumber\\
&& \times \left( \frac{z_{24}}{z_{14}} \right)^{h_{12}} \left( \frac{\bar z_{24}}{\bar z_{14}} \right)^{\bar h_{12}}
\left( \frac{z_{14}}{z_{13}} \right)^{h_{34}} \left( \frac{\bar z_{14}}{\bar z_{13}} \right)^{\bar h_{34}}
\sum_{\OO,\OO'} C_{12}^\OO g_{\OO \OO'} C_{34}^{\OO'}\, g^{1234}_{h,\bar h}(z,\bar z)
~,
\eea
where $z_{ij} = z_i-z_j$,
\beq
\label{mltheorybb}
g^{1234}_{h,\bar h}(z,\bar z) = z^h \bar z^{\bar h}  \, _2 F_1(h-h_{12},h+h_{34};2h;z)
\, _2 F_1(\bar h-\bar h_{12},\bar h+\bar h_{34};2 \bar h;\bar z)
\eeq
and
\beq
\label{mltheorybc}
z = \frac{z_{12}z_{34}}{z_{13}z_{24}} ~, ~~
\bar z = \frac{\bar z_{12}\bar z_{34}}{\bar z_{13}\bar z_{24}}
\eeq
the complex parameters $z,\bar z$ that express the cross-ratios $u,v$ in \eqref{mltheoryag}. We are also using the notation $h_{ij} = h_i - h_j$, while $_2 F_1(a,b;c;z)$ is the ordinary hypergeometric function. Adapting \eqref{mltheoryai}, we also set
\beq
\label{mltheorybd}
\sum_{\OO, \OO'\, | \, \Delta_\OO = \Delta_{\OO'} = h+\bar h} C_{12}^\OO g_{\OO \OO'} C_{34}^{\OO'} \equiv \, _s \C_{h,\bar h}
\eeq
suppressing the reference to the operators $\OO_i$.

In the above notation the crossing equations \eqref{mltheoryaj} take the form
\begin{align}
\label{mltheorybe}
&  \sum_{h,\bar h} {_s} \C_{h,\bar h}\, g^{(1234)}_{h,\bar h}(z,\bar z) =\cr
                                                                         &=(-1)^{(h_{41}+\bar h_{41})} \frac{z^{h_1+h_2}}{(z-1)^{h_2+h_3}} \frac{\bar z^{\bar h_1+\bar h_2}}{(\bar z-1)^{\bar h_2+\bar h_3} }
\sum_{h',\bar h'} {_t} \C_{h',\bar h'}\, g^{(3214)}_{h',\bar h'}(1-z,1-\bar z)\;.
\end{align}
At this point it is useful to make the following observations.

First, when one sums over the conformal block of a spinning quasi-primary operator (i.e.\ an operator with conformal weights $(h,\bar h)$  and $h\neq \bar h$) in either channel, one is also summing over a quasi-primary with conformal weights $(\bar h ,h)$. When we exchange $h$ and $\bar h$, the spin $s \to -s$, and the corresponding OPE-squared coefficients $\C_{h,\bar h}$ and $\C_{\bar h, h}$ are not in general equal. However, when the external operators are spinless, the OPE-squared coefficients are equal, $\C_{h,\bar h}=\C_{\bar h, h}$, and we can collect together the $(h,\bar h)$ and $(\bar h, h)$ contributions to form a single conformal block of the form
\begin{align}
\label{mltheorybf}
\tilde g^{(1234)}_{h,\bar h}(z,\bar z) = \frac{1}{1+\delta_{h,\bar h}} \bigg[
z^h \bar z^{\bar h}  & \, _2 F_1(h-h_{12},h+h_{34};2h;z)\cr
&\times \, _2 F_1(\bar h-\bar h_{12},\bar h+\bar h_{34};2 \bar h;\bar z) +(z \leftrightarrow \bar z) \bigg]
~.
\end{align}
In this manner, we can restrict the sums in \eqref{mltheorybe} to only run over operators with $h\geq \bar h$, hence reducing by half the number of intermediate quasi-primary operators that we need to consider in the ensuing application of the RL algorithm.

Second, it is useful to single-out the contribution of the identity operator, when this is present in a given channel, by setting $\C_{0,0}\, g_{0,0}^{(1234)}(z,\bar z) = g_{12} g_{34}$. This explicit non-vanishing constant in \eqref{mltheorybe} will prevent, in general, the RL algorithm from converging to the trivial solution where all $_s \C_{h,\bar h}$ and $_t \C_{h',\bar h'}$ are set to zero.

\subsection{Truncations, Spin-partitions and Measures of Accuracy}
\label{defspinpartitions}

We view the exact crossing equations \eqref{mltheorybe} as non-linear equations for the unknown positive\footnote{The positivity of the conformal weights $h,\bar h$ follows from well-known unitarity constraints in two dimensions.} conformal scaling dimensions $\Delta=h+\bar h$ and the corresponding OPE-squared coefficients $\C_{h,\bar h}$ in both channels. The spin $s=h-\bar h$ of the intermediate operators and the conformal weights $(h_i,\bar h_i)$ $(i=1,2,3,4)$ of the external operators are assumed to be given. However, in their current form, the {\it exact} crossing equations \eqref{mltheorybe} are impractical both for analytic and numerical methods. As already mentioned in Sec.~\ref{ingredients}, we need to implement a truncation.

For numerical methods the first obvious obstacle is the appearance of a typically infinite number of contributions to the conformal-block expansion. We address this problem by truncating the spectrum of intermediate quasi-primary operators, by setting some upper cutoff $\dmax$ on the scaling dimensions. The convergence properties of the conformal-block expansion \cite{Pappadopulo:2012jk} imply that one does not have to consider very large values of $\dmax$ for sensible numerical results, but the precise value of an optimal $\dmax$ is not easy to determine a priori and is, in general, theory-dependent. We will later make the surprising observation that in some examples values of $\dmax$ as low as 2 can already yield good approximations.\footnote{It may be that such behaviour is correlated with the fact that a CFT is easily truncable, in the sense of \cite{Gliozzi:2013ysa}. In general, however, truncability is not a pre-requisite for the application of our method.}

A second issue has to do with the continuous dependence of the exact crossing equations \eqref{mltheorybe} on the cross-ratio parameters $z,\bar z$. In this paper, we follow the approach of \cite{CastedoEcheverri:2016fxt} and evaluate the truncated crossing equations at a finite discrete set of points in the $z$-plane. We have noticed experimentally that the sampling of $z$-points suggested in Sec.~3.1 of \cite{CastedoEcheverri:2016fxt} works well also in our computations. In general, if the number of unknown scaling dimensions and OPE-squared coefficients is, in total, $N_{unknown}$, we choose $N_z$ $z$-points (with $N_z>N_{unknown}$) to evaluate the truncated crossing equations.

With these specifications, the exact crossing equations \eqref{mltheorybe} have been reduced to a finite set of non-linear algebraic equations, where the scaling dimensions of all contributing intermediate quasi-primary operators are bounded from above by $\dmax$. This necessarily also puts an upper bound on the allowed spin $s$ of these operators, since $|s|\leq \Delta \leq \dmax$.\footnote{Truncations on the spin of the conformal-block expansion and suitable discretisations in cross-ratio space are also commonplace in standard applications of the numerical conformal bootstrap.} However, despite the above considerable simplifications, the problem remains intractable: there is still a vast space of possibilities that an algorithm can explore associated with the freedom to choose any number of quasi-primaries at each spin. This final issue can be fixed by introducing a {\it spin-partition}.

The spin-partition is a sequence of positive integers that specifies the number of quasi-primaries per spin contributing to the conformal-block expansions of the truncated crossing equations. The spin-partition is an input to the RL algorithm that we set up in the next section. It fixes the dimensionality $N_{unknown}$ of the vector space of parameters $( \vec \Delta, \vec \C)$ where the search takes place. We will be listing spin-partitions using the template of Tab.~\ref{table:1}.

\begin{table}[t!]
\centering
\begin{tabular}{| c || c | c | c | c | c | c ||}
 \hline
 Spin &  0 & 1 & 2 & $\cdots$ & $n-1$ & $n$  \\ [0.5ex]
 \hline\hline
s-channel & $a_0$ & $a_1$ & $a_2$ & $\cdots$ & $a_{n-1}$ & $a_n$ \\
t-channel & $b_0$ & $b_1$ & $b_2$ &  $\cdots$ & $b_{n-1}$ & $b_n$ \\ [1ex]
 \hline
\end{tabular}
\caption{A depiction of the spin-partition for a truncated spectrum of integer-valued spins in a four-point function of spinless operators where the conformal-block expansions can be phrased in terms of only positive spins. In this example, we have chosen to use the same number of maximum spin in both $s$ and $t$ channels. The non-negative integers $a_i$, $b_i$ specify the number of operators with the corresponding spin, in the corresponding channel. For such a spin-partition the total number of unknowns in our problem is $N_{unknown}=2 \sum_{i=0}^n (a_i+b_i)$. For each unknown scaling dimension there is a corresponding unknown OPE-squared coefficient, hence the factor of 2 in this expression for $N_{unknown}$.}
\label{table:1}
\end{table}

We have thus arrived at a framework of truncated equations
\beq
\label{defsspinsaa}
\vec \E(\vec \Delta, \vec \C) = 0
~,
\eeq
where the dimension of the vector $(\vec \Delta, \vec \C)$ is $N_{unknown}$ and the dimension of the vector $\vec \E$ is $N_z$. Each entry $\E_i$ $(i=1,\ldots, N_z)$ of the vector $\vec \E$ contains the evaluation of the truncated version of Eq.~\eqref{mltheorybe} at one of the points $(z_i,\bar z_i)$ in our $z$-sampling
\bea
\label{defsspinsab}
&&\E_i = \sum_{h,\bar h}^{trunc} {_s} \C_{h,\bar h}\, g^{(1234)}_{h,\bar h}(z_i,\bar z_i)
\\
&&- (-1)^{(h_{41}+\bar h_{41})} z_i^{h_1+h_2} \bar z_i^{\bar h_1+\bar h_2} (z_i-1)^{-h_2-h_3} (\bar z_i-1)^{-\bar h_2-\bar h_3}
\sum_{h',\bar h'}^{trunc} {_t} \C_{h',\bar h'}\, g^{(3214)}_{h',\bar h'}(1-z_i,1-\bar z_i)
~,\nonumber
\eea
where $\displaystyle{\sum^{trunc}}$ denotes the truncated sum over intermediate operators.

This framework is very similar to the starting point of the approach \cite{Gliozzi:2013ysa,Li:2017ukc}. Notice, however, that the truncation in the scheme of \cite{Gliozzi:2013ysa,Li:2017ukc} is arbitrary, whereas here it comes with a further assumption that the unknown scaling dimensions are inside a specific window of scaling dimensions. This detail is an important distinction between our approach/implementation and those of \cite{Gliozzi:2013ysa,Li:2017ukc}. In particular, our approach entails a probabilistic search in specified parameter windows.

In general, \eqref{defsspinsaa} is not expected to have any exact solutions. Accordingly, as we explain in the next section, our RL algorithm is designed to minimise the Euclidean norm of $\vec \E$ and determine configurations of CFT data that satisfy the truncated crossing equations with the best possible accuracy. Although the Euclidean norm $|| \vec \E ||$ is an important quantity of the computation, it is not straightforward to judge whether its raw value at an optimal configuration is actually small or large. For that reason, we find it useful to define a ``relative measure of accuracy'', $\A$, defined in the context of \eqref{defsspinsab} as
\beq
\label{defsspinsac}
\A = \frac{|| \vec \E ||}{\E_{abs}}
\eeq
with
\bea
\label{defsspinsad}
&&\E_{abs} =\sum_{i=1}^{N_z} \Bigg[ \sum_{h,\bar h}^{trunc} \bigg | {_s} \C_{h,\bar h}\, g^{(1234)}_{h,\bar h}(z_i,\bar z_i) \bigg|
\\
&&+ \bigg| z_i^{h_1+h_2} \bar z_i^{\bar h_1+\bar h_2} (z_i-1)^{-h_2-h_3} (\bar z_i-1)^{-\bar h_2-\bar h_3} \bigg|
\sum_{h',\bar h'}^{trunc} \bigg| {_t} \C_{h',\bar h'}\, g^{(3214)}_{h',\bar h'}(1-z_i,1-\bar z_i) \bigg| \Bigg]
~.\nonumber
\eea
The quantity $\A$ is guaranteed to be a number between 0 and 1. Its value gives a \% measure of the accuracy at which we have been able to satisfy the truncated equations \eqref{defsspinsaa}, and this can in turn be compared more straightforwardly between different computations.

%%%%%%%%%%%%%%%%%%%%%%%%%%%%%%%%%%%%%

\section{Continuous Action Space Reinforcement Learning}
\label{ml}

In many physical settings it is very common to have access to large amounts of data (e.g.~collider physics), where supervised/unsupervised ML techniques find direct application. However, in scenarios often found in theoretical physics this is not usually the case. This is where RL comes in handy because the learning agent is able to generate its own data.

Reinforcement Learning, in brief, is an algorithm consisting of two parts with equal importance. The first is the so-called ``agent'', which is the brain of the algorithm. The second is the ``environment'': what the agent interacts with. The basic setup of the algorithm is the process of the agent making decisions as it explores the provided environment, while the environment gives feedback on the agent's actions. One wants the agent to explore the environment towards finding an ideal solution, while exploiting the best solution it finds (explore-exploit dilemma). One also has to find a suitable algorithm for how the agent (the neural network) ``learns'' and retains its experiences.

There exists a considerable amount of previous work on DRL algorithms, which have been applied to a large variety of problems, both theoretical and real-world. There are examples of agents which can beat video games, drive cars, guide robots, solve mathematical equations and---possibly the most famous one---AlphaGo, which beat professional Go champions using a combination of supervised learning and DRL \cite{AlphaGo,Gopaper}, and the improved AlphaGo Zero, which relied completely on DRL \cite{AlphaGoZero}.

Such algorithms can be split into two main sets and can be distinguished by whether the actions (defined by numbers) taken by the agent are discrete or continuous. Algorithms such as Deep Q-Learning \cite{Mnih:2015jgp} or Actor-Critic methods \cite{DBLP:journals/corr/abs-2102-04376} use a discrete action space (convenient when one can take only a finite amount of actions), while algorithms such as the soft Actor-Critic method \cite{DBLP:journals/corr/abs-1801-01290} and the Deep Deterministic Policy Gradient method \cite{Lillicrap2016ContinuousCW} were developed for when the actions can take any real value.

In this paper we are making use of the soft Actor-Critic algorithm and implementing it using the PyTorch package for Python 3.7, but one could have equivalently chosen the Deep Deterministic Policy Gradient or any of the other Machine Learning libraries (TensorFlow etc.). We will not go into the details of the aforementioned algorithms, since these can be found in the original papers (with pseudo code), and there exist plenty of additional online resources showcasing their implementation. Furthermore, we will treat the learning algorithm itself as a black box, i.e.\ we will not be interested in its study, although one can adjust the hyperparameters for the learning following \cite{DBLP:journals/corr/abs-1801-01290}. We are mostly interested in the environment the agent will get to explore.

\subsection{Soft Actor-Critic Algorithm}

Although we will not be providing the full details of the agent implementation, it is still useful to give a short overview of what actually happens inside the brain of the algorithm.

The algorithm itself is an iterative process, where the iteration is over ``steps'' taken by the agent. These steps can also be grouped into ``episodes''. An episode is concluded when the last step results in a terminal state. The steps and terminal states are more important when talking about the environment, and they will be discussed in more detail in the following subsection. In every iterative step of the algorithm there are a number of processes executed by the code. In order, these include:
\begin{enumerate}
  \item {\it Choose Action}: Since our agent is designed to come up with scaling dimensions and OPE-squared coefficients for given CFT spin-partitions, each action will directly correspond to an unknown (such as a scaling dimension or OPE-squared coefficient). The actions themselves can take continuous values. The agent takes an action by predicting values for the unknowns.
  \item {\it Implement the Action in the Environment}: We shall explain the implementation of the environment in detail in the next subsection. For now we shall say that the values of the predictions by the agent are fed into the environment code.
  \item {\it Observe the Environment}: In this step the constraints are calculated by the environment (such as the crossing equations or additional constraints) and are fed back to the agent as observations (it is what the agent ``sees'').
  \item {\it Obtain Reward}: The algorithm for the environment comes up with a quantitative judgment (discussed in the next section) on how well the agent did with its prediction of the parameters. This is then fed back to the agent.
  \item {\it Check if Final State}: The environment simply checks if the agent managed to predict something which has a better reward than the previous best. This tells the agent to try and find better solutions.
  \item {\it Update Memory Buffer}: In the algorithm we use previous agent experiences (i.e. previous steps) that are stored in what is called an experience replay buffer: a multidimensional array containing all the information fed to the agent from previous iterations. This is very important for the next step. In the current step the current information is stored in the array.
  \item {\it Update Neural Networks (learn)}: A random set of samples is taken from the previously mentioned memory buffer and this data is used as training data to update the weights of the neural networks of the learning algorithm. In the optimisation step of the weights we use the ADAM optimiser \cite{Kingma2015AdamAM}. Once the networks have been fed forward and backpropagated, their structures (weights) will adjust to better suit the data. Hence in the next iteration they will try to predict results which better satisfy the constraints. It is important to note that the networks do not actually predict the values themselves but a probability distribution which is then sampled for the predictions; this is where the explore-exploit dilemma enters.
\end{enumerate}

We display the details of the NNs that we used for our searches in Tab.~\ref{table:hyperparameters}.

\begin{table}[t!]
    \centering
     \begin{tabular}{ | c | c |}
     \hline
      NN Hyperparameter &  Value \\ \hline\hline
      learning rates &        0.0005 \\ \hline
      $\gamma$ (discount factor) & 0.99\\  \hline
      replay buffer size & 100000  \\  \hline
      batch size & 64  \\ \hline
      $\tau$ (smoothing coefficient)	&	  0.001\\  \hline
      layer 1 size & 128  \\  \hline
      layer 2 size & 64 \\  \hline
      reward scale	& 	0.005	\\ 
        \hline
    \end{tabular}
    \caption{Hyperparameter values for the NNs used in our calculations, presented in the format of \cite{DBLP:journals/corr/abs-1801-01290}.}
    \label{table:hyperparameters}
    \end{table}

\subsection{Environment}

Here we summarise some of the most salient features of the environment implementation. The latter guides the agent's learning on how to predict the CFT data. Since implementations of RL agents can be easily adapted for use in a large variety of problems, setting the environment becomes the most important part of the implementation. The environment must provide an interface that the agent can interact with, calculate the constraints, come up with a quantitative notion of success and define a terminal state.

The environment in which our agent ``moves'' is the space of parameters $(\vec \Delta, \vec \C)$. Every value for the scaling dimensions/OPE-squared coefficients defines a different theory. For our purposes, a point $(\vec \Delta, \vec \C)$ in parameter space is judged based on how well it satisfies the numerical constraints of truncated crossing equations $\vec \E (\vec \Delta, \vec \C) = 0$, \eqref{defsspinsaa}.

The agent's predictions feed into these numerical constraints. Since we have truncated the equations and are numerically approximating the values (and the number of constraints is larger than the number of unknowns) it is unlikely that there will be a solution that exactly satisfies all constraints in \eqref{defsspinsaa}. In fact, one ends up with deviations from zero for each constraint, which then have to be minimised so that the constraints are satisfied to as good a numerical approximation as possible. These deviations are individual numbers that form the observations of the agent.

One can now straightforwardly define the reward function. Clearly, the agent should be encouraged to pick values for the parameters which minimise all the constraints. The simplest choice for such a reward is
\begin{align}
  \label{eq:1}
R := -||\vec \E||  
\end{align}
The use of the Euclidean norm of the vector $\vec \E$ is natural (but not unique) as it quantifies the distance from the origin where the truncated equations \eqref{defsspinsaa} are satisfied exactly. The negative sign punishes larger distances away from the origin more than smaller ones. It would be interesting in the future to further explore how the efficiency of the algorithm depends on the choice of reward and to examine other options, e.g. the possibility of different weights in the definition of the Euclidean norm.

The very last section of the environment checks for final states. In our case this is simply a flag checking if the current solution is better than the current best from previous runs. If, indeed, it is, then the code overwrites the previous best, and supplies the flag to the agent. The agent needs to know whether or not the step led to a final state, as this directly feeds into the approximation of the probability distribution.

We summarise these steps in Alg.~\ref{alg:code0}, where $A$ stands for an  action by the agent and $R^*$ for the current best reward.

\subsection{Three Modes of Running the Algorithm}
\label{modes}

The RL algorithm can be implemented in several different ways depending on the scope and focus of the search. In this subsection, we outline three different modes that were employed in producing the results of Secs~\ref{minimal} and \ref{boson}. In summary, these are:

\begin{itemize}

\item {\bf Mode 1.} Specify the spin-partition and $\dmax$ and search for scaling dimensions within the unitarity bound and $\dmax$. For OPE-squared coefficients there are very few constraints, e.g.\ they may only be restricted by unitarity to be positive.
  
\item {\bf Mode 2.} There is a specific expectation for the scaling dimensions, for which the search is contained within a narrow window. There are no expectations for the OPE-squared coefficients, where the search is initially as wide as in mode 1.

\item {\bf Mode 3.} Both scaling dimensions and OPE-squared coefficients are within a specified, known narrow window. This mode could be implemented as a supplementary run after a mode 1 or mode 2 run, or it could be relevant in cases where we are verifying an analytic solution in the context of the truncated crossing equations, or in cases where the solution is known in some regime of parameters and we are changing these parameters adiabatically.

\end{itemize}

\vspace{.25cm}
\begin{algorithm}[H]
  \label{alg:code0}
  \footnotesize{
    \SetAlgoLined
\KwIn{$A$, $R^*$}
\KwOut{individual constraints, $R$, $R^*$}

 Env calculate constraints using $A$\;
  Env calculate $R$\;
  Env check if $R > R^*$\;
  Agent observe individual constraints\;
  Agent store memory in buffer\;
  Agent learn\;
  \If{$R > R^*$}{
   overwrite previous best reward $R^*=R$\;
 }}
 \caption{Basic Reinforcement-Learning Routine}
\end{algorithm}
\vspace{.5cm}

Clearly, the range of the search becomes more narrow as we go from mode 1 to mode 3. The computational time is expected to be larger, in general, in mode 1.

Our algorithm gives the user two key dials that can be tuned at will at the beginning, or multiple times in the middle of a run. The first is a lower bound for each parameter (we will call it the ``floor''). The second dial is a separate size for the search window of each parameter, in each action of the agent (we will call this dial the ``guess-size''). As a rule of thumb, the initial window should at first be set large enough to minimise the probability of the agent getting trapped at a local minimum. Once the presence of a potential global minimum has been established, one can then start to hone in by gradually reducing its size. We next provide a more detailed description of each mode.

\subsubsection{Mode 1}

Since this mode involves the widest search windows, a blind search may be hindered by the existence of multiple false vacua, or may lead to an approximate solution that represents a CFT that is not of immediate interest. As a result, this mode can be assisted by additional preparation that partially restricts the search. For example, one could start with a rough preliminary exploration of the minima of $||\vec \E||$ using Mathematica, or obtain a rough estimate of some of the scaling dimensions using the approach of \cite{Gliozzi:2013ysa}. This preparation can help significantly facilitate the subsequent search.

To commence the search we initially run the algorithm in ``guessing mode'' where the RL agent only tries to improve on its own guess in the current cycle. This allows for the random exploration of configuration space and generates some initial profiles of CFT data. 

Then, we enter the ``normal mode'', where the agent initially takes the final state from the guessing mode and tries to find small corrections so as to better satisfy the constraints. Once it finds such a correction, it replaces the final state and proceeds with a new correction iteratively. Here one can set specific values for the floor and guess-sizes. It helps to set the guess-size at a magnitude comparable to the expected order of parameter change as the agent hits the next final state. In most cases, the user can easily detect this size by observing how the agent generates configurations in real time.

The algorithm continues the search ad infinitum and the crucial question is when to stop and record the result. We have observed in the context of different theories that in actual solutions the agent reaches in reasonable time (of the order of an hour on a modern laptop) a value of the relative measure of accuracy $\A$ below 0.5\%. In addition, when the search window is set near actual solutions the agent keeps reducing $\A$ significantly below the threshold of 0.5\% with an apparent convergence on the values of the parameters $(\vec \Delta, \vec \C)$. Based on this observation, we have always aimed for runs that drop $\A$ below 0.1\%.

\subsubsection{Mode 2} In this mode we conduct, from the beginning, a narrow search in scaling dimensions. We have found that the following protocol produces good results.

We set the floor of the scaling dimensions to the expected values and the corresponding guess-sizes to 0. This freezes the scaling dimensions and reduces the dimensionality of the search by half, since we are conducting a search by varying only the OPE-squared coefficients. After exiting the guessing mode, we conduct the search for the optimal OPE-squared coefficients using the same procedure as in mode 1.

Once the relative accuracy $\A$ drops to the order of 1\%, we unfreeze the scaling dimensions by reducing their floor and opening their guess-size. The size of the search window around the expected values of the scaling dimensions can be controlled freely by the user. If the agent is already in the vicinity of a solution, the scaling dimensions will not move significantly once unfrozen, and the full set of parameters $(\vec \Delta, \vec \C)$ will now be adjusted by the agent to reduce $\A$ even further. We continue the search until we achieve an acceptably small value of $\A$ and observe an apparent convergence following the general procedure outlined in mode 1.

During this process it may happen that some scaling dimensions are driven towards the boundary of the prescribed window of search. In that case, the user can slightly increase the corresponding window to explore whether the approximate solution lies nearby. As long as the agent keeps improving the accuracy $\A$, the window can be kept in place. If there is, however, a stage in the run where the agent stops improving at an unacceptably high $\A$, and the adjustment of guess-sizes does not help, then this can be viewed as a strong signal that a solution does not exist in the prescribed windows.

\subsubsection{Mode 3} In this case, we are conducting a narrow search in all components of the parameters $(\vec \Delta, \vec \C)$. We can run the algorithm as in mode 2 without the initial run to approximate the configuration of the OPE-squared coefficients, since this is already approximately known.

\subsubsection{Enlarging the Spin-Partition}\label{sec:enlarge}

After having obtained results for a given spin-partition one can implement a shortcut for subsequent searches with an enlarged spin-partition (e.g. when $\Delta_{\rm max}$ is increased). Instead of re-running the algorithm for all parameters, it is more economical to instead implement a strategy akin to that of mode 2:

\begin{itemize}
    \item Perform the search with the least number of parameters using the steps outlined previously.
  \item Freeze these parameters.

  \end{itemize}
 \vspace{.1cm}

\begin{algorithm}[H]
  \label{alg:code}
  \footnotesize{
    \SetAlgoLined
\KwIn{spin partition, floor, guess-size}
\KwOut{$(\vec \Delta, \vec \C)$}
 initialise Agent (memory buffer $+$ NN weights)\;
 initialise file for overall best reward $R^*$\;
 \While{running guessing mode}{
  Agent choose action\;
  Env calculate constraints\;
  Env calculate $R$\;
  Env check if $R > R^*$\;
  Agent observe current state\;
  Agent store memory in buffer\;
  Agent learn\;
  \If{$R > R^*$}{
   overwrite previous best result, $R^* = R$\;
   }
 }
 \While{not accurate enough}{
  reinitialise Agent (memory buffer $+$ NN weights)\;
  \While{running normal mode}{
   Agent choose action\;
   Env calculate constraints\;
   Env calculate $R$\;
   Env check if $R > R^*$\;
   Agent observe current state\;
   Agent store memory in buffer\;
   Agent learn\;
   \If{$R > R^*$}{
    overwrite previous best result, $R^* = R$\;
    }
   \If{Agent trapped}{
    break normal mode loop\;
    }
  }
 }
 \If{adding new parameters}{
  rerun above code first freezing then unfreezing\;
  }}
 \caption{Reinforcement-Learning CFT Data Search}
\end{algorithm}
\newpage

  \begin{itemize}

  \item Start adding the new dynamical parameters to the set of frozen ones to approximately find
    the new global minimum.

  \item Unfreeze all parameters and let the agent determine how these new
    parameters change the old ones to find a better solution.
\end{itemize}

This type of implementation opens up the exciting possibility of reconstructing considerable amounts of CFT data without a full, specific, a priori given spin-partition.

\subsubsection{Comments on User Input}\label{sec:freezing}

To summarise, our overall approach is sketched in Alg.~\ref{alg:code}. It should be apparent from the description of the above three modes that, although the RL algorithm is set up to run independently without the input of an external user, in actual runs user intervention can help in significantly speeding up the search. A suitable real-time adjustment of the guess-size for individual parameters helps the agent focus faster around a region of potential interest. In the future, this is an aspect of the algorithm we would like to improve---or better automate---in order to facilitate more efficient parallel runs. At this stage, the mode with the minimal user input is mode 3, which involves the smallest search windows.

%%%%%%%%%%%%%%%%%%%%%% 

\section{Application I: Minimal Models}
\label{minimal}

We now pass on to explicit applications of our algorithm, starting with minimal models. The unitary minimal models are, in the appropriate sense, the simplest possible 2D CFTs and benchmarks of the original conformal bootstrap programme from the 1970s. Here we revisit them from the perspective of the global part of the Virasoro algebra, completely disregarding the Virasoro enhancement of the $so(2,2)$ conformal algebras.

In this section we search for approximate solutions to the crossing equations that we listed in Sec.~\ref{2dintro}, which describe minimal models. The consistency of the crossing equations in this well-known class of 2D CFTs was understood analytically early on. It is therefore a good starting point to verify that our method recovers known facts about these theories correctly. We focus on the two leading representatives in the series of unitary minimal models, the Ising and tri-critical Ising models.

\subsection{Analytic Solution}

We next briefly recall some of the salient features of the Ising and tri-critical Ising models (see \cite{DiFrancesco:1997nk} for a comprehensive review).

\subsubsection{Ising Model}

The Ising model, $\MM(4,3)$, is the simplest model in the unitary minimal series $\MM(p+1,p)$. It has central charge $c=\frac{1}{2}$ and it is equivalent to the CFT of a free Majorana fermion. Besides the identity operator $\II$, its spectrum contains two more primary operators: the spin operator $\sigma$ with conformal weights $(h,\bar h)=(\frac{1}{16},\frac{1}{16})$, and the energy-density operator (also called thermal operator) $\varepsilon$ with conformal weights $(h,\bar h) = (\frac{1}{2},\frac{1}{2})$. The corresponding OPEs are
\bea
\label{minaa1}
&&\sigma \times \sigma = [\II ] + [\varepsilon ]
\\
\label{minaa2}
&&\sigma \times \varepsilon = [\sigma]
\\
\label{minaa3}
&&\varepsilon \times \varepsilon = [\II ]
~,
\eea
where $[\OO]$ denotes the Virasoro conformal family of the primary $\OO$. In what follows, we will study the four-point functions
\begin{align}
  \label{eq:5}
&  \langle \sigma(z_1,\bar z_1) \sigma(z_2,\bar z_2) \sigma(z_3,\bar z_3) \sigma(z_4,\bar z_4)\rangle~,\\
& \langle \varepsilon(z_1,\bar z_1) \varepsilon(z_2,\bar z_2) \varepsilon(z_3,\bar z_3) \varepsilon(z_4,\bar z_4)\rangle~.\label{eq5:2}
\end{align}
The conformal-block decomposition of these correlation functions contains, according to the first and third OPEs in \eqref{minaa1}, \eqref{minaa3}, the quasi-primaries in the Virasoro conformal family of the identity and energy-density operators. By definition, a quasi-primary state (in the holomorphic sector) is annihilated by the $L_1=\frac{1}{2\pi i} \oint dz\, z^2 T(z)$ conformal generator. Equivalently, the OPE between the energy-momentum tensor $T(z)$ and a quasi-primary should have no $z^{-3}$ pole. It is straightforward to construct these quasi-primaries by acting on the primary state with the Virasoro raising operators $L_{-k}$, $(k\geq 1)$ but one needs to take into account the structure of the Virasoro algebra and the presence of null states in the corresponding Verma modules. States of the form $L_{-1} |{\rm state}\rangle$ are, by definition, descendants in the sense of the $so(2,2)$ global part of the conformal algebra.

For example, by focusing on the holomorphic part of the theory, we obtain at the first few levels the following quasi-primaries in the Virasoro conformal families of the identity and energy-density operators.\footnote{This computation is greatly facilitated by the Mathematica package {\bf FeynCalc9.3.1} \cite{Mertig:1990an,Shtabovenko:2016sxi,Shtabovenko:2020gxv}.} In the conformal family of the identity, the states
\beq
\label{minab}
L_{-2} |0\rangle ~, ~~ \left( L_{-2}^2 - \frac{3}{10} L_{-1} L_{-3} \right)|0\rangle, ~~
\left( L_{-2} L_{-3} - \frac{1}{2} L_{-1} L_{-2}^2 - \frac{1}{6} L_{-1} L_{-4} \right) |0\rangle
\eeq
are the only quasi-primaries up to level 5. In the conformal family of the energy-density, the states
\beq
\label{minaca}
|\varepsilon \rangle ~, ~~ \left( L_{-3} -\frac{4}{9} L_{-1} L_{-2} \right) |\varepsilon \rangle ~, ~ ~
\left( L_{-4} +\frac{10}{27} L_{-2}^2 - \frac{5}{9} L_{-1} L_{-3}  \right) |\varepsilon \rangle
~,\nonumber
\eeq
\beq
\label{minacb}
\left( L_{-5} -\frac{2}{3} L_{-1} L_{-4} +\frac{5}{24} L_{-1}^2 L_{-3} - \frac{1}{40} L_{-1}^5  \right) |\varepsilon \rangle
\eeq
are the only quasi-primaries up to level 5. A potential quasi-primary at level 2 does not exist, because it is one of the characteristic null states of the Ising model.

When combined with the anti-holomorphic sector, these results yield the spin-partitions that will be employed in the analysis of Sec.~\ref{ising} below.

\subsubsection{Tri-critical Ising Model}

The tri-critical Ising model, $\MM(5,4)$, is the next minimal model in the unitary series.\footnote{One of the beautiful features of the tri-critical Ising model is that it is secretly endowed with supersymmetry \cite{Friedan:1984rv}, but this feature will not play any role in our analysis.} It has central charge $c=\frac{7}{10}$, and besides the identity operator, its conformal primary spectrum comprises three energy-density operators
\bea
\label{minad}
&&\varepsilon ~~{\rm with}~~ (h,\bar h) = \left( \frac{1}{10}, \frac{1}{10} \right)~, \nonumber\\
&&\varepsilon' ~~{\rm with}~~ (h,\bar h) = \left( \frac{3}{5}, \frac{3}{5} \right)~, \nonumber\\
&&\varepsilon'' ~~{\rm with}~~ (h,\bar h) = \left( \frac{3}{2}, \frac{3}{2} \right)~, \nonumber
\eea
and two spin operators
\bea
\label{minad}
&&\sigma ~~{\rm with}~~ (h,\bar h) = \left( \frac{3}{80}, \frac{3}{80} \right)~, \nonumber\\
&&\sigma' ~~{\rm with}~~ (h,\bar h) = \left( \frac{7}{16}, \frac{7}{16} \right)~. \nonumber
\eea
The OPEs of these operators are listed in Tab.~7.4 of \cite{DiFrancesco:1997nk}. We will be interested in four-point functions of the tri-critical Ising model that resemble those of the Ising model, and the way our algorithm differentiates between the two CFTs. We will therefore focus on the primary operators $\sigma'$ and $\varepsilon''$, which satisfy
\beq
\label{minae}
\sigma' \times \sigma' = [\II]+[\varepsilon'']~, ~~ \varepsilon'' \times \varepsilon'' = [\II]
~.
\eeq
Notice the similarity with the OPEs \eqref{minaa1}, \eqref{minaa3}. Accordingly,  in the next subsection we will study the four-point functions
\begin{align}
  \label{eq:4}
  &\langle \sigma'(z_1,\bar z_1) \sigma'(z_2,\bar z_2) \sigma'(z_3,\bar z_3) \sigma'(z_4,\bar z_4)\rangle~,\\
    &  \langle \varepsilon''(z_1,\bar z_1) \varepsilon''(z_2,\bar z_2) \varepsilon''(z_3,\bar z_3) \varepsilon''(z_4,\bar z_4)\rangle~.\label{eq4:2}
\end{align}
Similar to the case of the Ising-model primary $\varepsilon$, we find that the conformal family of $\varepsilon''$ in the tri-critical Ising model contains the following quasi-primary states, up to level 4 in the holomorphic sector:
\beq
\label{minafa}
\left( L_{-2} - \frac{3}{8} L_{-1}^2 \right) | \varepsilon'' \rangle~, ~~
\left( L_{-2}^2 + \frac{43}{2240} L_{-1}^4 -\frac{15}{56} L_{-1}^2 L_{-2} \right) | \varepsilon'' \rangle~,
\nonumber
\eeq
\beq
\label{minafb}
\left( L_{-4} + \frac{31}{672} L_{-1}^4 - \frac{5}{28} L_{-1}^2 L_{-2} \right) |\varepsilon'' \rangle
~.
\eeq
To obtain this result we had to use that the Verma module of the state $|\varepsilon''\rangle$ contains the following null state at level 3 (in the holomorphic sector):
\beq
\label{minag}
\left( L_{-3} - \frac{4}{7} L_{-1} L_{-2} + \frac{4}{35} L_{-1}^3 \right) |\varepsilon'' \rangle
~.
\eeq

\subsection{Reinforcement-Learning  Results}

The above analytic data can now be compared with those obtained from our RL algorithms. This exercise is helpful in checking the efficiency of our code before proceeding to the more complicated example of the $c=1$ compactified boson CFT.

\subsubsection{$\langle \sigma \sigma \sigma \sigma\rangle$ in Ising Model}
\label{ising}

The exact crossing equation for the four-point function \eqref{eq:5} in the Ising model is
\beq
\label{isingaa}
{\sum_{h \geq \bar h}}' \C_{h,\bar h} \Big( |z-1|^{2\Delta_\sigma} \tilde g_{h,\bar h}^{(\sigma \sigma \sigma \sigma)}(z,\bar z) - |z|^{2\Delta_\sigma}  \tilde g_{h,\bar h}^{(\sigma \sigma \sigma \sigma)}(1-z,1-\bar z) \Big)
+|z-1|^{2\Delta_\sigma} - |z|^{2\Delta_\sigma}=0
~.
\eeq
As this correlator involves four identical spinless operators, both channels, $s$ and $t$, exchange the same intermediate operators with even spin. In the last two terms we have singled out the contribution of the identity operator and hence the sum ${\sum}'$ does not contain it.

\begin{table}[t!]
\centering
\begin{tabular}{ | c || c | c | c | c | c | c | c ||}
 \hline
	Spin 		& 0 	& 1 	& 2 	& 3	& 4	& 5	& 6	\\ [0.5ex]
 \hline\hline
			& 2 	& - 	& 1 	& - 	& 1	& - 	& 1 \\ [1ex]
 \hline
\end{tabular}
\caption{A spin-partition informed by the conformal block decomposition of the four-point function $\langle\sigma(z_1,\bar z_1) \sigma(z_2,\bar z_2) \sigma(z_3,\bar z_3) \sigma(z_4,\bar z_4)\rangle$ in the Ising model with $\dmax=6.5$.}
\label{table:Ising_spin_partitions}
\end{table}

Using the crossing equation \eqref{isingaa} to determine our reward function, we performed the following computation with the RL algorithm. We set $\Delta_\sigma = \frac{1}{8}$, for the external spin operator $\sigma$, and searched in mode 2 for solutions with the spin-partition of Tab.~\ref{table:Ising_spin_partitions}, which is informed by the analytic solution with a cutoff $\dmax = 6.5$. A more agnostic search in mode 1, with more limited information about the initial profile of the scaling dimensions, is also feasible. Such runs are presented in the next Sec.~\ref{boson}. Here, the mode-2 runs are computing independently the OPE-squared coefficients and confirm the analytic values of the scaling dimensions that were used to initiate the runs. In the implementation of the algorithm we enforced the unitarity constraint that the OPE-squared coefficients are positive. 

This is a search in a 10-dimensional space of unknowns (5 for the scaling dimensions and 5 for the corresponding OPE-squared coefficients). The results of a run with 29 crossing equations---that is, \eqref{isingaa} evaluated at 29 different points on the $z$-plane---appear in Tab.~\ref{table:sigma_results}. This particular run took approximately 12 hours on a modern laptop machine to yield the relative accuracy $\A = 3.31618 \times 10^{-6}$.\footnote{After submission to PRD, we found that replacing the MPMATH numerical PYTHON library with SciPy reduces the running time to just 30 minutes with similar results to those of Tab.~\ref{table:sigma_results}.} When unfrozen, the scaling dimensions were allowed to vary with a guess-size 0.1. It is worth noting that the agent started the run with a random profile of OPE-squared coefficients (some of which were orders of magnitude away from those of the Ising model) and gradually converged to the results of Tab.~\ref{table:sigma_results}.

\begin{table}[t!]
\centering
\begin{tabular}{ | c || c | c || c | c |}
 \hline
\multicolumn{5}{|c|}{$\Delta_\sigma=\frac{1}{8}$ } \\
 \hline\hline
 spin 	& analytic $\Delta$ 	& RL $\Delta$ 	& analytic $\C$  	& RL $\C$ 				\\ [0.5ex]
 \hline\hline
  0		& 4 				& 3.9331603	& 2.44141$ \times 10^{-4}$		& 3.657538$ \times 10^{-4}$			\\
  0		& 1 				& 0.9881525	& 0.25			& 0.25254947				\\
  2		& 2 				& 1.9802496	& 0.015625		& 0.015717817				\\
  4		& 4 				& 3.9497		& 2.19727$ \times 10^{-4}$		& 2.4715587 $ \times 10^{-4}$			\\
  6		& 6 				& 5.971367	& 1.36239$ \times 10^{-5}$ 	& $0.54007314 \times 10^{-5}$	\\ \hline \hline
\multicolumn{5}{|c|}{$\A = 3.31618 \times 10^{-6}$} \\ \hline
\end{tabular}
\caption{Analytic and numerical solutions for scaling dimensions and OPE-squared coefficients in the conformal-block decomposition of the four-point function $\langle \sigma(z_1,\bar z_1) \sigma(z_2,\bar z_2) \sigma(z_3,\bar z_3) \sigma(z_4,\bar z_4)\rangle$ for $\Delta_\sigma=\frac{1}{8}$ and the spin-partition of Tab.~\ref{table:Ising_spin_partitions} with $\dmax =6.5$. The numerical results were obtained with a mode-2 run of the RL algorithm.}
\label{table:sigma_results}
\end{table}

We observe that the relative accuracy at which we can satisfy the truncated crossing equations is impressively strong, even with a very rough truncation of only 5 quasi-primary operators. When compared against the analytic expectations, the numerical results for the scaling dimensions agree at the order of 1\%. For the OPE-squared coefficients, the agreement is equally impressive for the two lower-lying operators $\varepsilon$ and $L_{-2}$ with scaling dimensions 1 and 2 respectively, but (as might be expected) becomes worse for the higher scaling dimension operators at $\Delta=4,6$ that lie closer to $\dmax$.

Notice that the exact unitarity bound for the spin-2, 4 and 6 operators requires their scaling dimensions satisfying $\Delta \geq 2,4$ and 6 respectively. Since we have truncated the crossing equations, we do not expect the results to obey the strict unitarity bounds, and, as a result, we have allowed the agent to explore solutions with a small violation of these bounds.

\subsubsection{$\langle \sigma' \sigma' \sigma' \sigma' \rangle$ in Tri-critical Ising Model}
\label{triising}

Similarly, in the tri-critical Ising model we study the four-point function \eqref{eq:4} whose crossing equation is
\beq
\label{triisingaa}
{\sum_{h \geq \bar h}}' \C_{h,\bar h} \Big( |z-1|^{2\Delta_{\sigma'}} \tilde g_{h,\bar h}^{(\sigma' \sigma' \sigma' \sigma')}(z,\bar z) - z^{2\Delta_{\sigma'}}  \tilde g_{h,\bar h}^{(\sigma' \sigma' \sigma' \sigma')}(1-z,1-\bar z) \Big)
+|z-1|^{2\Delta_{\sigma'}} - |z|^{2\Delta_{\sigma'}} =0
~.
\eeq
Once again, the sum over $h,\bar h$ does not include the contribution of the identity operator, which has been singled out in the last two terms of the equation. In this case we ran the RL algorithm in mode 2 by setting $\Delta_{\sigma'}=\frac{7}{8}$ for the external operator $\sigma'$, and using the spin-partition of Tab.~\ref{table:triIsing_spin_partitions} informed by the analytic solution of the tri-critical Ising model with $\dmax = 6.5$.

\begin{table}[t!]
\centering
\begin{tabular}{ | c || c | c | c | c | c | c | c ||}
 \hline
	Spin 		& 0 	& 1 	& 2 	& 3	& 4	& 5	& 6	 \\ [0.5ex]
 \hline\hline
			& 2 	& - 	& 3 	& -	& 1	& -	& 1	 \\ [1ex]
 \hline
\end{tabular}
\caption{A spin-partition informed by the conformal-block decomposition of the four-point function $\langle \sigma'(z_1,\bar z_1) \sigma'(z_2,\bar z_2) \sigma'(z_3,\bar z_3) \sigma'(z_4,\bar z_4)\rangle$ in the tri-critical Ising model with $\dmax= 6.5$.}
\label{table:triIsing_spin_partitions}
\end{table}

 It may be instructive to compare this spin-partition with the corresponding spin-partition for the Ising model in Tab.~\ref{table:Ising_spin_partitions}. The only difference is 3 versus 1 spin-2 quasi-primary operators. In the analytic solution there is another difference, which is not apparent in Tab.~\ref{table:triIsing_spin_partitions}. At spin-6 the tri-critical Ising model has 2 degenerate quasi-primary states
\beq
\label{triisingab}
\left( L_{-2}^3 + \frac{10}{7} L_{-6} -\frac{1}{2} L_{-1} L_{-2} L_{-3} \right) |0\rangle~, ~~
\left( L_{-3}^2 + \frac{92}{63} L_{-6} -\frac{4}{9} L_{-1} L_{-2} L_{-3} \right) |0\rangle
~,
\eeq
instead of just one, whose contribution combines as a single term in the crossing equations. The degeneracies are, therefore, invisible to the spin-partition and consequently not detectable from our analysis.

\begin{table}[t!]
\centering
\begin{tabular}{ | c || c | c || c | c | c |}
 \hline
\multicolumn{5}{|c|}{$\Delta_{\sigma'}=\frac{7}{8}$} \\
 \hline \hline
 spin 	& analytic $\Delta$ 	& RL $\Delta$ 	& analytic $\C$  & RL $\C$ 		\\ [0.5ex]
 \hline\hline
  0		& 4 				& 3.8950076	& 0.299072	& 0.63403654		\\
  0		& 3 				& 2.9969018	& 0.285171	& 0.29550505		\\
  2		& 2 				& 1.97196		& 0.546875	& 0.6054145		\\
  2		& 6 				& 5.97496		& 0.0238323	& 0.041339442		\\
  2		& 5 				& 5.0424104	& 0.0270531	& 0.040516548		\\
  4		& 4 				& 4.051943	& 0.0435791	& 0.06928008		\\
  6		& 6 				& 5.9997706	& 0.00589177	& 0.0047707544	\\ \hline \hline
 \multicolumn{5}{|c|}{$\A = 0.000705966$} \\ \hline
\end{tabular}
\caption{Analytic and numerical solutions for scaling dimensions and OPE-squared coefficients in the conformal-block decomposition of the four-point function $\langle \sigma'(z_1,\bar z_1) \sigma'(z_2,\bar z_2) \sigma'(z_3,\bar z_3) \sigma'(z_4,\bar z_4)\rangle$ for $\Delta_\sigma=\frac{7}{8}$ and the spin-partition of Tab.~\ref{table:triIsing_spin_partitions} with $\dmax=6.5$. The numerical results were obtained with a mode-2 run of the RL algorithm.}
\label{table:sigmap_results}
\end{table}

In this context, we performed a search in a 14-dimensional space of scaling dimensions and OPE-squared coefficients. The RL algorithm was run with 29 different points on the $z$-plane. The results that appear in Tab.~\ref{table:sigmap_results} were obtained after a run that lasted approximately 8 hours and yielded a configuration with relative accuracy $\A = 0.000705966$ (significantly larger than that in Tab.~\ref{table:sigma_results} for the Ising model).

The comparison between the numerical and analytic results follows a pattern similar to that in the Ising model. The agent has clearly located the CFT data of the tri-critical Ising model, and the agreement with the analytic results is better for the low-lying operators at spin-0 and spin-2 with expected scaling dimensions 3 and 2 respectively.

\subsubsection{4-point Functions with the Identity as the Single Virasoro Conformal Block}

Several minimal models have 4-point functions of a single conformal primary with the identity as the only Virasoro conformal block. In the Ising model, $\MM(4,3)$, this feature appears in the four-point function $\langle \varepsilon(z_1,\bar z_1) \varepsilon(z_2,\bar z_2) \varepsilon(z_3,\bar z_3) \varepsilon(z_4,\bar z_4)\rangle$, in the tri-critical Ising model, $\MM(5,4)$, in the four-point function $\langle \varepsilon''(z_1,\bar z_1) \varepsilon''(z_2,\bar z_2) \varepsilon''(z_3,\bar z_3) \varepsilon''(z_4,\bar z_4)\rangle$, in the three-state Potts model, $\MM(6,5)$, in the four-point function $\langle Y(z_1,\bar z_1) Y(z_2,\bar z_2) Y(z_3,\bar z_3) Y(z_4,\bar z_4) \rangle$ etc. The operators $\varepsilon, \varepsilon'', Y$ are all spinless with different scaling dimensions: $1,3,6$, respectively. In this subsection, we compare the first two cases: $\langle \varepsilon \varepsilon \varepsilon \varepsilon\rangle$ in the Ising model, and $\langle \varepsilon'' \varepsilon'' \varepsilon''  \varepsilon'' \rangle$ in the tri-critical Ising model.

In all these cases the crossing equations are similar,
\beq
\label{idblockaa}
{\sum_{h \geq \bar h}}' \C_{h,\bar h} \Big( |z-1|^{2\Delta_\OO} \tilde g_{h,\bar h}^{(\OO \OO \OO \OO)}(z,\bar z) - z^{2\Delta_\OO}  \tilde g_{h,\bar h}^{(\OO \OO \OO \OO)}(1-z,1-\bar z) \Big)
+|z-1|^{2\Delta_{\OO}} - |z|^{2\Delta_{\OO}} =0
~,
\eeq
and the spin-partition is the same. $\OO$ is the spinless external operator and $\Delta_\OO$ its scaling dimension.

Using the spin-partition of Tab.~\ref{table:id_spin_partitions}, which contains the expected number of quasi-primary operators in the identity Virasoro block up to scaling dimension 6.5, we varied the scaling dimension $\Delta_\OO$ of the external operator and searched for solutions to the crossing equations \eqref{idblockaa}. Our main purpose in this subsection was to verify the expected analytic solutions of the Ising and tri-critical Ising models and that the algorithm could distinguish solutions with different external scaling dimensions but the same spin-partition. For these purposes a mode-3 run was deemed sufficient.

\begin{table}[t!]
\centering
\begin{tabular}{ | c || c | c | c | c | c | c | c |}
 \hline
	Spin 		& 0 	& 1 	& 2 	& 3	& 4 	& 5	& 6	 \\ [0.5ex]
 \hline\hline
			& 1 	& - 	& 2 	& -	& 1	& - 	& 1	 \\ [1ex]
 \hline
\end{tabular}
\caption{A spin-partition for the conformal block contribution of the identity operator with $\dmax= 6$.}
\label{table:id_spin_partitions}
\end{table}

The results of Tab.~\ref{table:epsilon_results} were obtained with $\OO=\varepsilon$. Indeed, they verify quite clearly the expected structure of the Ising model. The run reported in Tab.~\ref{table:epsilon_results} took only 2 hours to reach the relative accuracy $\A = 0.000862723$ in mode 3.

\begin{table}[t!]
\centering
\begin{tabular}{ | c || c | c || c | c |}
 \hline
\multicolumn{5}{|c|}{$\Delta_\varepsilon = 1$ } \\
 \hline\hline
 spin 	& analytic $\Delta$ 	& RL $\Delta$ 	& analytic $\C$  	& RL $\C$ 			\\ [0.5ex]
 \hline\hline
  0		& 4 				& 4.0683885	& 1				& 1.0427935			\\
  2		& 2 				& 1.9544389	& 1				& 1.1926383			\\
  2		& 6 				& 5.926708	& 0.1				& 0.1150967			\\
  4		& 4 				& 3.904911	& 0.1				& 0.20634486			\\
  6		& 6 				& 5.9300733	& 0.0238095	 	& 0.022085898			\\ \hline \hline
\multicolumn{5}{|c|}{$\A = 0.000862723$} \\ \hline
\end{tabular}
\caption{Analytic and numerical solutions for scaling dimensions and OPE-squared coefficients in the conformal block decomposition of the four-point function $\langle \varepsilon(z_1,\bar z_1) \varepsilon(z_2,\bar z_2) \varepsilon(z_3,\bar z_3) \varepsilon(z_4,\bar z_4)\rangle$ for $\Delta_\varepsilon = 1$ and the spin-partition of Tab.~\ref{table:id_spin_partitions} with $\dmax =6.5$. The numerical results were obtained with a mode-3 run of the RL algorithm.}
\label{table:epsilon_results}
\end{table}

A similar mode-3 run with $\OO=\varepsilon''$ produced the results of Tab.~\ref{table:epsilonPP_results} with a comparable relative accuracy $\A = 0.000668002$. The general features of the expected structure of the tri-critical Ising model are present, but some of the numbers (depicted in magenta in Tab.~\ref{table:epsilonPP_results}) exhibit significant discrepancies with the analytic results. A possibly related feature in the analytic solution is the presence of sizeable OPE-squared coefficients at higher scaling dimensions. In order to probe this feature further, we repeated the computation with a higher cutoff, $\dmax = 8.5$, which involves a spin-partition with 8 different operators. The resulting 16-dimensional search in mode 3 produced the numbers listed in Tab.~\ref{table:epsilonPP_results_8}, which exhibit a definite improvement compared to the previous $\dmax = 6.5$ run. For the convenience of the reader we have highlighted with a magenta color the corresponding numbers in Tabs~ \ref{table:epsilonPP_results} and \ref{table:epsilonPP_results_8}.

\begin{table}[t!]
\centering
\begin{tabular}{ | c || c | c || c | c |}
 \hline
\multicolumn{5}{|c|}{$\Delta_{\varepsilon''} = 3$ } \\
 \hline\hline
 spin 	& analytic $\Delta$ 	& RL $\Delta$ 			& analytic $\C$  	& RL $\C$ 				\\ [0.5ex]
 \hline\hline
  0		& 4 				& {\color{magenta}5.3342843}	& 41.3265			& 43.009876				\\
  2		& 2 				& {\color{magenta}2.586108}	& 6.42857			& 6.317041				\\
  2		& 6 				& 5.900023			& 23.4184			& 23.202938				\\
  4		& 4 				& {\color{magenta}4.8769903}	& 3.64286			& {\color{magenta}16.4788}		\\
  6		& 6 				& 6.0306115			& 1.23387		 	& 1.7063767				\\ \hline \hline
\multicolumn{5}{|c|}{$\A = 0.000668002$} \\ \hline
\end{tabular}
\caption{Analytic and numerical solutions for scaling dimensions and OPE-squared coefficients in the conformal block decomposition of the four-point function $\langle \varepsilon''(z_1,\bar z_1) \varepsilon''(z_2,\bar z_2) \varepsilon''(z_3,\bar z_3) \varepsilon''(z_4,\bar z_4)\rangle$ for $\Delta_{\varepsilon''} = 3$ and the spin-partition of Tab.~\ref{table:id_spin_partitions} with $\dmax =6.5$. The numerical results were obtained with a mode-3 run of the RL algorithm.}
\label{table:epsilonPP_results}
\end{table}

\begin{table}[t!]
\centering
\begin{tabular}{ | c || c | c || c | c |}
 \hline
\multicolumn{5}{|c|}{$\Delta_{\varepsilon''} = 3$ } \\
 \hline\hline
 spin 	& analytic $\Delta$ 	& RL $\Delta$ 				& analytic $\C$  	& RL $\C$ 				\\ [0.5ex]
 \hline\hline
  0		& 4 				& {\color{magenta}4.505229}	& 41.3265			& 40.726093				\\
  0		& 8 				& 7.9896655				& 13.2704			& 13.0988035				\\
  2		& 2 				& {\color{magenta}2.3935893}	& 6.42857			& 5.4763665				\\
  2		& 6 				& 7.1316943				& 23.4184			& 21.824356				\\
  4		& 4 				& {\color{magenta}4.362866}	& 3.64286			& {\color{magenta}4.9283843}	\\
  4		& 8 				& 7.9502306				& 5.89678			& 6.0844507				\\
  6		& 6 				& 6.0996165				& 1.23387			& 2.7852516				\\
  8		& 8 				& 8.006613				& 0.251744	 	& 0.0013012796			\\ \hline \hline
\multicolumn{5}{|c|}{$\A = 0.000771919$} \\ \hline
\end{tabular}
\caption{A $\dmax = 8.5$ version of Tab.~\ref{table:epsilonPP_results}.}
\label{table:epsilonPP_results_8}
\end{table}

In the above computations we fixed the scaling dimension of the external operator and tried to determine the remaining data. It would be interesting to perform a more general computation, where the scaling dimension of the external operator is one of the unknowns of the search. In mode 1, this search should be able to identify, solely from the input of the spin-partition, different solutions corresponding to the data of each CFT in the minimal series. We do not perform this computation here, but present results of a very similar computation in Sec.~\ref{spin1} in the case of the compactified boson CFT.

%%%%%%%%%%%%%%%%%%%%%%%%%%%%%%%%

\section{Application II: $c=1$ Compactified Boson}
\label{boson}

With an eye towards more general applications, it is important to explore the performance of our approach beyond the restricted class of rational conformal field theories, of which minimal models are a special case. In this section, we study the $c=1$ compactified boson CFT. This is a free scalar CFT. Free CFTs are the benchmark of the Lagrangian QFT approach and the basis of perturbative methods in quantum field theory, readily solved by traditional methods and an entry-level litmus test for the generalisation of our method to more challenging settings. 

The reader should appreciate that by rediscovering the compactified boson CFT as a solution to the crossing equations, one would be able to solve it without the use of the standard Lagrangian methods, e.g.\ they would be able to determine correlation functions without using Wick's theorem. Despite its simplicity, the free scalar CFT has a rich spectrum of primary operators with momentum and winding around the target circle and scaling dimensions that depend non-trivially on an exactly marginal coupling---the radius of the circle. This is therefore an interesting toy model where our methods can be used to compute non-trivial CFT data across a continuous family of CFTs connected by exactly marginal deformations, namely across a conformal manifold. Conformal manifolds are ubiquitous in four-dimensional supersymmetric QFTs, e.g.\ in 4D $\NN=4$ SYM theory, which would be one of the natural subsequent applications of the RL approach presented here.

We study two examples of four-point functions in the compactified boson CFT: four-point functions of vertex operators with momentum or winding, and four-point functions of the conserved $U(1)$ current. We discover that even with a very small cutoff, as low as $\dmax=2$, the algorithm can detect correctly the 2D compactified boson CFT and returns rather accurate approximate values for scaling dimensions and OPE-squared coefficients.

\subsection{Analytic Solution}
\label{s1anal}

Before delving into the results of the RL exercise, it is useful to recall briefly the analytic solution of the 2D $S^1$ scalar theory that we want to rediscover from a conformal bootstrap/RL perspective.

Consider the 2D CFT of a compact boson $X$ with radius $R$:
\beq
\label{s1analaa}
S = \frac{1}{4\pi} \int d^2 z \d X \bar \d X~, ~~ X \simeq X + 2\pi R
~.
\eeq
Since this is a free theory, it is straightforward to analytically compute all its data. Let us summarise some of the pertinent details following closely the conventions of \cite{Polchinski:1998rq} with $\alpha'=2$.

The basic conformal primaries of the theory are the $U(1)$ currents
\beq
\label{s1analac}
j(z) = \frac{i}{2} \d X(z)~, ~~ \bar j(\bar z) = \frac{i}{2} \d X(\bar z)
\eeq
and the vertex operators\footnote{All the operators appearing below should be understood as being normal ordered.}
\beq
\label{s1analad}
V_{p,\bar p} (z,\bar z) = e^{ipX(z) + i \bar p \bar X(\bar z)}~, ~~ p=\frac{n}{R}+\frac{wR}{2}~, ~~ \bar p = \frac{n}{R}-\frac{wR}{2}
~,
\eeq
where $n$ and $w$ are the integer momentum and winding quantum numbers of the corresponding states. $j$, $\bar j$ have respectively conformal scaling dimensions $(h,\bar h) = (1,0), (0,1)$, while $V_{p,\bar p}$ has $(h,\bar h) = (\frac{p^2}{2},\frac{\bar p^2}{2})$. The spin of an operator is $s=h-\bar h$. As a result, the vertex operator $V_{p,\bar p}$ has spin $s=\frac{1}{2}(p^2-\bar p^2) = nw$. Corresponding states with only momentum, or only winding, are spinless.

The remaining spectrum of operators can be organised using the Virasoro algebra, but since we only want to use the global $so(2,2)$ part of the 2D conformal algebra, we need to also identify all the quasi-primary operators. All quasi-primaries of the theory can be obtained by combining any quasi-primary operator from the left-moving (holomorphic) sector with any quasi-primary operator from the right-moving (anti-holomorphic) sector. There are no factors with mixed holomorphic-antiholomorphic derivatives in an operator because of the equations of motion $\d \bar \d X=0$. Hence, let us focus momentarily on the holomorphic sector.

As already noted in our minimal-model discussion, a quasi-primary state (in the holomorphic sector) is annihilated by the $L_1=\frac{1}{2\pi i} \oint dz\, z^2 T(z)$ conformal generator. This requires that the OPE between the energy-momentum tensor $T(z)$ and a quasi-primary should have no $z^{-3}$ pole. The general vertex operator with holomorphic momentum $p$ has the form
\beq
\label{s1analae}
\OO_{m_1,\ldots,m_r;p} \equiv \prod_{a=1}^r (\d^a X)^{m_a} \, e^{ip X}
~.
\eeq
A straightforward computation shows that the $z^{-3}$ pole in the OPE $T(z) \OO_{m_1,\ldots,m_r;p}(0)$ is
\beq
\label{s1analaf}
\Big[ T(z) \OO_{m_1,\ldots,m_r;p}(0) \Big]_3 = \sum_{a=2}^r (a-1)a \, \OO_{m_1,\ldots, m_{a-1}+1, m_a-1, m_{a+1},\ldots,m_r;p}(0)
~.
\eeq
A generic quasi-primary is a linear combination of operators of the form \eqref{s1analae} with the same conformal dimension. Eq.~\eqref{s1analaf} can be used to determine the numerical coefficients in these combinations. For example, the quasi-primaries with up to six derivatives are:
\bea
\label{s1analag}
&&\left[ (\d X)^2 + i p \d^2 X \right] \, e^{ip X}~, ~~ \left[ (\d X)^3 + \frac{3}{2} i p \d X \d^2 X - \frac{p^2}{4} \d^3 X \right] \, e^{ip X}~,
\nonumber\\
&& \left[ (\d X)^4 + 2 i p (\d X)^2 \d^2 X - p^2 (\d^2 X)^2 \right] \, e^{ip X}~, ~~ \left[ \d X \d^3 X + \frac{i p}{12}\d^4 X - \frac{3}{2} (\d^2 X)^2 \right] \, e^{ip X}~, ~~
\nonumber\\
&& \left[ (\d X)^5 + \frac{5 i p}{2} (\d X)^3 \d^2 X - \frac{p^2}{4} \left( (\d X)^2 \d^3 X + i p \d^2 X \d^3 X + 6 \d X (\d^2 X)^2 \right) \right] \, e^{ip X}~, ~~
\nonumber\\
&& \left[ (\d X)^2 \d^3 X - \frac{3}{2} \d X (\d^2 X)^2 +\frac{p }{4} \left( \frac{5 i }{6} \d X \d^4 X - \frac{ 1}{24} \d^5 X - i \d^2 X \d^3 X\right) \right] \, e^{ip X}~,~~
\nonumber
\\
&& \left[ (\d X)^6 + 3i p (\d X)^4 \d^2 X - 3 p^2 \left( (\d X)^2 (\d^2 X)^2 + \frac{ip}{3}(\d^2 X)^3 \right) \right] \, e^{ip X}~, ~~
\nonumber\\
&& \bigg[ (\d X)^3 \d^3 X -\frac{3}{2} \left( (\d X)^2 (\d^2 X)^2 + \frac{i p}{3} (\d^2 X)^3 \right)
\nonumber\\
&&\hspace{5cm} +\frac{3 ip}{2} \left( \d X \d^2 X \d^3 X - (\d^2 X)^3 + \frac{ip}{12} (\d^3 X)^2 \right) \bigg] \, e^{ip X}~, ~~
\nonumber\\
&& \left[ \d X \d^5 X + \frac{ip}{30} \d^6 X - 10 \d^2 X \d^4 X + 10 (\d^3 X)^2 \right] \, e^{ip X}~,~~
\nonumber\\
&& 	\bigg[ (\partial X)^3 \partial^3 X - \frac{3}{2} (\partial X)^2 (\partial^2 X)^2 + \frac{3 p^2}{8} (\partial^3 X)^2 - \frac{ p^2}{3} \partial^2 X \partial^4 X ~~
\nonumber\\
&&\hspace{5cm}  + \frac{i p}{3} (\partial X)^2 \partial^4 X - \frac{i p}{2} \partial X \partial^2 X \partial^3 X \bigg] e^{i p X}  ~.
\eea

Putting together the holomorphic and anti-holomorphic parts, general quasi-primaries can be obtained as linear combinations of the operators
\beq
\label{s1analai}
\OO_{\{m_a\},\{ \bar m_{\bar a}\};p,\bar p} \equiv \prod_{a=1}^r \left( \d^a X \right)^{m_a} \prod_{\bar a=1}^{\bar r} \left( \d^{\bar a} \bar X \right)^{\bar m_{\bar a}}  e^{ipX+i\bar p \bar X}
~.
\eeq
The conformal dimensions of these operators are
\bea
\label{s1analaj}
&&h = \ell + \frac{1}{2} p^2 =  \ell + \frac{1}{2} \left( \frac{n}{R} + \frac{w R}{2} \right)^2
~,
\\
&& \bar h = \bar \ell + \frac{1}{2} \bar p^2 = \bar \ell + \frac{1}{2} \left( \frac{n}{R} - \frac{w R}{2} \right)^2\;,
\eea
where $\ell = \displaystyle{\sum_{a=1}^r} a \, m_a$, $\bar \ell = \displaystyle{\sum_{\bar a=1}^{\bar r}} \bar a \, \bar m_{\bar a}$ and $p, \bar p$ are expressed in terms of the momentum and winding quantum numbers.

The two- and three-point functions involving the above quasi-primaries can be computed straightforwardly using Wick contractions. Explicit results, that will be compared against those from the RL output, will be listed in the next subsection.

\subsection{Reinforcement-Learning Results}
\label{s1}

We will now attempt to rediscover the $S^1$ theory from the conformal-bootstrap perspective. We consider two kinds of four-point functions. The first one is the four-point function of four spinless conformal primaries with arbitrary, but fixed, scaling dimension. The zero-spin assumption is not necessary; we only make it here for convenience and illustration purposes. We further assume that these operators are charged under a conserved $U(1)$ symmetry. We denote them as $V_p$ and parametrise their scaling dimension $\Delta_p$ by the real variable $p$ using the relation
\beq
\label{s1aa}
\Delta_p \equiv p^2
~.
\eeq
We emphasise that this equation should be viewed as the definition of the real number $p$. At this point we do not specify how $p$ relates to the $U(1)$ charge of $V_p$ and hence \eqref{s1aa} is {\it not} a dynamical statement about the scaling dimension $\Delta_p$ in terms of some other quantum number.

Keeping the above in mind, we consider the four-point function
\beq
\label{s1ab}
\langle V_p(z_1,\bar z_1) V_p(z_2,\bar z_2) \bar V_p(z_3,\bar z_3) \bar V_p(z_4,\bar z_4) \rangle
~,
\eeq
where $\bar V_p$ denotes the complex conjugate of $V_p$. Since $V_p$ and $\bar V_p$ have opposite $U(1)$ charge, the four-point function \eqref{s1ab} is neutral under the assumed global $U(1)$ symmetry. $V_p$ is expected to capture the primary vertex operator $V_{p,p}(z,\bar z)= e^{i p (X(z) + \bar X(\bar z))}$ with $p = \bar p = \frac{n}{R}$ and winding $w=0$, or the T-dual $V_{p,-p}(z,\bar z) = e^{i p (X(z) - \bar X(\bar z))}$ with $p = \bar p = \frac{w}{R}$ and momentum $n=0$. Only a minimal part of this information will be incorporated indirectly into the algorithm via the spin-partition. Using this partial information, the agent will have to uncover that $V_p$ is indeed part of the $S^1$ theory and that $p$ is related to the $U(1)$ charge.

The second kind of four-point function that we will consider is the correlator of the conserved spin-1 operator $j$,
\beq
\label{s1ac}
\langle j(z_1) j(z_2) j(z_3) j(z_4) \rangle
~.
\eeq
We next display the results of the RL algorithm for each case.

\subsubsection{Momentum/winding Sector}
\label{momentumsector}

\begin{table}[t!]
\centering
\begin{tabular}{ | c || c || c | c | c | c | c | c |}
 \hline
			& Spin 		&  0 	& 1 	& 2 	& 3 	& 4 	& 5  \\ [0.5ex]
 \hline\hline
$\dmax = 2$ 	& s-channel 	& 1 	& -- 	& -- 	& --	& -- 	& --  \\
			& t-channel 	& 1 	& 1 	& 1 	& -- 	& -- 	& --  \\ [1ex]
 \hline \hline
 $\dmax = 3.5$ 	& s-channel 	& 1 	& -- 	& 1 	& 1 	& --	& --  \\
			& t-channel 	& 1 	& 2 	& 1 	& 1 	& -- 	& --  \\ [1ex]
 \hline \hline
 $\dmax = 4.5$ 	& s-channel 	& 2 	& -- 	& 1 	& 1 	& 1 	& --  \\
			& t-channel 	& 2 	& 2 	& 2 	& 1 & 1 & --  \\ [1ex]
 \hline \hline
 $\dmax = 5.5$ 	& s-channel 	& 2 	& 1	& 1 	& 1 	& 1 	& 1  \\
			& t-channel 	& 2 	& 3 	& 2 	& 2 	& 1 	& 1  \\ [1ex]
 \hline
\end{tabular}
\caption{Spin-partitions for the conformal-block decomposition of the four-point function $\langle V_p(z_1,\bar z_1) V_p(z_2,\bar z_2) \bar V_p(z_3,\bar z_3) \bar V_p(z_4,\bar z_4) \rangle$ at four different values of the cutoff $\dmax$.}
\label{table:Mom_spin_partitions}
\end{table}

The crossing equation for the four-point function \eqref{s1ab} can be written as
\beq
\label{s1bb}
\sum_{h\geq \bar h} {_s} \C_{h,\bar h} |z-1|^{2\Delta_p} \tilde g_{h,\bar h}^{(VV\bar V \bar V)}(z,\bar z)
-  {\sum_{h'\geq \bar h' }}'
  {_t} \C_{h',\bar h'} |z|^{2\Delta_p} \tilde g_{h,\bar h}^{(\bar V V V \bar V)}(1-z,1-\bar z)
- |z|^{2 \Delta_p} = 0
~.
\eeq
In the $t$-channel block decomposition we have separated the contribution of the identity operator and have used the normalisation convention $\langle V_p \, \bar V_p \rangle =1$.

Let us fix for concreteness the scaling dimension $\Delta_p$ of $V_p$ to some specific value, e.g.\ $\Delta_p=0.1$. This value is deliberately small to allow for spin-partitions with relatively small cutoff $\dmax$. In Tab.~\ref{table:Mom_spin_partitions} we collect four spin-partitions that will be used to study the truncated version of the crossing equations \eqref{s1bb}. These spin-partitions are inspired by the analytic solution of the $S^1$ theory when imposing the cutoff $\dmax = 2,3.5,4.5,5.5$, respectively, in the OPEs of the $s$- and $t$-channels. In each of these spin-partitions the number of unknowns (scaling dimensions plus OPE-squared coefficients) that we are solving for is 8, 16, 26, 36.

In Tabs~\ref{table:Mom_results_1}-\ref{table:Mom_results_3} we have collected the expected analytic results of the $S^1$ theory for $V_p = V_{p,\pm p}$ and $|p|=\sqrt{0.1}$ together with the best results of the runs we performed. In contrast to Sec.~\ref{minimal}, where we presented results based mainly on mode-2 and mode-3 runs (guided by partial prior information about the CFT data in the initialisation of the code), in this section we present results of genuine mode-1 runs based only on the information provided by the spin-partition.

One of the first observations in Tab.~\ref{table:Mom_results_1} is that already in the simplest case of $\Delta_{\rm max}=2$ the RL algorithm predicts the corresponding CFT data to very good accuracy. The run reported in Tab.~\ref{table:Mom_results_1} for $\dmax=2$ used 30 $z$-points and took approximately 2 hours to yield the relative accuracy $\A=0.000197442$. The results for the higher cutoffs, that incorporate further operators with higher conformal scaling dimensions (and spin), were obtained by building on the $\dmax=2$ data with the use of the incremental mode-1 procedure of Sec.~\ref{sec:enlarge}. 

The results at $\dmax=3.5$ in Tab.~\ref{table:Mom_results_1} exhibit a noticeable decrease in $\A$ (which translates to a smaller violation of the truncated-reduced crossing equations) and agreement between the numerical and analytic results for the low-lying spectrum, which is comparable with the $\dmax=2$ run. Notice that there are two deliberate features complicating the $\dmax=3.5$ run. First, the fact that the spin-3 operator is absent in the $s$-channel was not an input. The agent had to discover this feature (as it does), but this complicates the search. Interestingly, although the spin-3 operator is absent in the exact conformal decomposition, the agent manages to identify its scaling dimension with remarkable accuracy. Apparently, this is not an accident; similar results are obtained in the higher cutoff runs of Tabs~\ref{table:Mom_results_2}-\ref{table:Mom_results_3}. Second, in the runs of Tab.~\ref{table:Mom_results_1} we are not using any information about the signs of the OPE-squared coefficients. As a result, some of the OPE-squared coefficients obtained in the $\dmax=3.5$ run have the wrong sign in the $t$-channel. Once again, this complicates the search and prevents the agent from improving the agreement between the numerical and analytic results.

\begin{table}[t!]
\centering
\begin{tabular}{ | c | c || c | c || c | c |}
 \hline
\multicolumn{6}{|c|}{$\dmax=2$} \\
\hline
 Channel 			& spin 	& analytic $\Delta$ 	& RL $\Delta$ 	& analytic $\C$  & RL $\C$ 	\\ [0.5ex]
 \hline\hline
$s$ 				& 0 		& 0.4 			& 	0.38694087		& 1			&	0.99479413		\\ [1ex]
 \hline \hline
$t$			 	& 0		 & 2 				& 2.1108415			& 0.01	& 0.010378244			\\
				& 1		 & 1 				& 	0.9485743		& -0.1		& -0.10135128			\\
 				& 2		 & 2 				& 	2.1295118		&0.005		&0.004827103			\\ \hline \hline
\multicolumn{6}{|c|}{$\A = 0.000197442$} \\ \hline
\end{tabular}

\vspace{1cm}

\begin{tabular}{ | c | c || c | c || c | c |}
 \hline
\multicolumn{6}{|c|}{$\dmax=3.5$} \\
 \hline
 Channel 	& spin 	& analytic $\Delta$ 	& RL $\Delta$ 	& analytic $\C$  & RL $\C$ 	\\ [0.5ex]
 \hline\hline
$s$ 		& 0 		& 0.4 			&  0.39011472			& 1			&	0.999143		\\
 		& 2 		& 2.4 			& 	2.2029796	 	& 3.57143 $\times 10^{-3}$	& 2.2229333 $\times 10^{-3}$		\\
  & 3 		& 3.4 	&	3.203875	& 	0 	& 4.971186 $\times 10^{-7}$				\\[1ex]
 \hline \hline
$t$  		& 0		 & 2 				& 2.1141205			& 0.01		& 0.008170344			\\
		& 1		 & 1 				& 0.95283717			& -0.1		& -0.09884554			\\
		& 1		 & 3 				& 2.8024354			& -5  $\times 10^{-4}$		& 9.701283 $\times 10^{-4}$			\\
 		& 2		 & 2 				& 2.1266346			& 0.005		&	0.003557264		\\
		& 3		 & 3 				& 2.8005629			& -1.66667  $\times 10^{-4}$	& 4.3958283  $\times 10^{-4}$			\\ \hline \hline
\multicolumn{6}{|c|}{$\A = 0.00000225745$} \\ \hline
\end{tabular}

\caption{Analytic and numerical solutions for scaling dimensions and OPE-squared coefficients for $\Delta_p=0.1$ and spin-partitions with $\dmax = 2,3.5$ and 30 $z$-points respectively. The numerical results were obtained using the mode described in Sec.~\ref{sec:enlarge}.}
\label{table:Mom_results_1}
\end{table}

The data reported in Tabs~\ref{table:Mom_results_2}-\ref{table:Mom_results_3} are based on multi-dimensional searches with an even larger number of operators (13 and 18 respectively). To increase the accuracy (namely, reduce the value of $\A$) we used the results of the incremental mode-1 procedure of Sec.~\ref{sec:enlarge} to initialise an additional, subsequent mode-2 run with 49 $z$-points. The mode-2 run began with a search on the OPE-squared coefficients alone, while the scaling dimensions were kept fixed at the values obtained from the prior mode-1 search. At a second stage of the run, the scaling dimensions were unfrozen and the agent was allowed to search in the complete space of scaling dimensions and OPE-squared coefficients to find the results reported in Tabs~\ref{table:Mom_results_2}-\ref{table:Mom_results_3}. In the $\dmax=4.5$ run we kept the signs of the OPE-squared coefficients free (as in Tabs~\ref{table:Mom_results_1}). With the exception of the OPE-squared coefficients for the second spin-0 operator in the $s$-channel, the agent managed to predict the correct signs. To illustrate what happens when we input the correct signs, we performed the more complicated $\dmax=5.5$ run by fixing the signs of the OPE-squared coefficients at their expected analytic values. The combined mode-1 and mode-2 runs at $\dmax=4.5$ took approximately 2 days and the runs at $\dmax=5.5$ 4 days.

\begin{table}[t!]
\centering
%\vspace{-0.4cm}
\begin{tabular}{ | c | c || c | c || c | c |}
 \hline
\multicolumn{6}{|c|}{$\dmax=4.5$} \\
 \hline
 Channel 	& spin 	& analytic $\Delta$ 	& RL $\Delta$ 	& analytic $\C$  & RL $\C$ 	\\ [0ex]
 \hline\hline
  $s$ 	& 0 		& 0.4 			&  0.4185128	& 1			&	0.98406625 \\
   		& 0 		& 4.4 			& 4.229518	& 1.27551 $\times 10^{-5}$	& -5.3023865 $\times 10^{-5}$	\\
 		& 2 		& 2.4 			& 2.4269097	& 3.57143 $\times 10^{-3}$	& 4.041962 $\times 10^{-3}$	\\
		& 3 		& 3.4 			& 3.2022634	& 0			&  -1.0026526 $\times 10^{-3}$ 	\\
                & 4 		& 4.4 			& 4.574162	& 1.96039 $\times 10^{-3}$	& 2.7667696 $\times 10^{-4}$ 			\\ [0.4ex]
 \hline \hline
$t$  		& 0		 & 2 				& 2.0097528	& 0.01		& 0.0025764485	\\
		& 0		 & 4 				& 3.8530886	& 2.5 $\times 10^{-5}$		& 4.1462967$\times 10^{-4}$		\\
		& 1		 & 1 				& 0.9313935	& -0.1		& -0.10908633	\\
		& 1		 & 3 				& 2.9478629	& -5 $\times 10^{-4}$			& -7.262531 $\times 10^{-3}$		\\
 		& 2		 & 2 				& 2.0496795	& 0.005					&	0.013589153	\\
		& 2		 & 4 				& 3.8056073	& 1.66667 $\times10^{-5}$	& 6.19941 $\times10^{-5}$			\\
		& 3		 & 3 				& 2.9541698    &  -1.66667 $\times10^{-4}$	& - 4.592793 $\times10^{-3}$		\\
		& 4		 & 4 				& 4.0146556    & 4.16667 $\times 10^{-6}$ 	& 4.924626 $\times 10^{-3}$		\\ \hline \hline
\multicolumn{6}{|c|}{$\A = 0.0000206548$} \\ \hline
\end{tabular}

%\vspace{-0.2cm}
\caption{Analytic and numerical solutions for scaling dimensions and OPE-squared coefficients for $\Delta_p=0.1$ and spin-partitions with $\dmax = 4.5$ and 49 $z$-points. The numerical results were obtained using a mode-2 run on top of the mode described in Sec.~\ref{sec:enlarge}.}
\label{table:Mom_results_2}
\end{table}

\begin{table}[t!h!]
\centering
\begin{tabular}{ | c | c || c | c || c | c |}
 \hline
\multicolumn{6}{|c|}{$\dmax=5.5$} \\
 \hline
 Channel 	& spin 	& analytic $\Delta$ 	& RL $\Delta$  &analytic $\C$ 	 	& RL $\C$ 	\\ [0ex]
 \hline\hline
$s$ 		& 0 		& 0.4 	&  0.40006787		& 1			&	1.0057276		\\
		& 0 		& 4.4 	& 4.336432		& 1.27551 $\times 10^{-5}$	& 0.43016876 $\times 10^{-5}$		\\
  		& 1 		& 5.4 	& 5.307818		& 0			& -2.2633198 $\times 10^{-4}$		\\
 		& 2 		& 2.4 	& 2.4060674		&  3.57143 $\times 10^{-3}$	& 5.486169 $\times 10^{-3}$ 		\\
		& 3 		& 3.4 	& 3.446559		& 0			& -0.4480493  $ \times 10^{-5}$	\\
		& 4 		& 4.4 	& 4.410344		& 1.96039 $\times 10^{-3}$	&	0.27796367 $\times 10^{-3}$	\\
  		& 5 		& 5.4 	& 5.3354797		& 0  			& -9.976282 $\times 10^{-5}$		\\[0.4ex]
 \hline \hline
$t$  		& 0		 & 2 		& 2.001293		& 0.01		& 0.0056684865	\\
		& 0		 & 4 		& 4.0166564		& 2.5 $\times 10^{-5}$		& 4.8836926 $\times 10^{-4}$	\\
		& 1		 & 1 		& 1.040068		& -0.1		& -0.085237		\\
		& 1		 & 3 		& 3.0494268		& -5 $\times 10^{-4}$			& -2.271628 $\times 10^{-2}$	\\
		& 1		 & 5 		& 4.9848695		& -8.33333 $\times 10^{-7}$  	& -9.268466 $\times 10^{-4}$	\\
 		& 2		 & 2 		& 2.00707			& 0.005		& 0.0018059064	\\
		& 2		 & 4 		& 4.045016		& 1.66667 $\times 10^{-5}$	& 7.282457 $\times 10^{-4}$	\\
		& 3		 & 3 		& 3.0331514		& -1.66667 $\times10^{-4}$	& -2.894943 $\times10^{-4}$	\\
		& 3		 & 5 		& 4.9544168		& -4.16667 $\times 10^{-7}$ 	& -3.3044117$\times 10^{-3}$	\\
		& 4		 & 4 		& 3.9395354		&  4.16667 $\times 10^{-6}$	& 6.668457 $\times 10^{-4}$	\\
		& 5		 & 5 		& 5.0390368		&  -8.33333 $\times 10^{-8}$	& -4.3607014 $\times 10^{-4}$	\\ \hline \hline
\multicolumn{6}{|c|}{$\A = 0.0000321653$} \\ \hline
\end{tabular}

\caption{Analytic and numerical solutions for scaling dimensions and OPE-squared coefficients for $\Delta_p=0.1$ and spin-partitions with $\dmax = 5.5$ and 49 $z$-points. The numerical results were obtained using a mode-2 run on top of the mode described in Sec.~\ref{sec:enlarge}.}
\label{table:Mom_results_3}
\end{table}

Comparing the numerical and analytic results in Tabs~\ref{table:Mom_results_2}-\ref{table:Mom_results_3} we observe that the agent has performed impressively well for the scaling dimensions (even for the odd-spin operators that do not contribute to the $s$-channel in the exact result). It performed decently for the OPE-squared coefficients of the low-lying $\dmax=2$ operators, but poorly for many of the remaining, numerically smaller coefficients. From the single runs reported in Tabs~\ref{table:Mom_results_2}-\ref{table:Mom_results_3} we can immediately deduce that the algorithm works, because it managed to minimise the violation of the truncated crossing equations and identified CFT data with a very low value of $\A$. To obtain a better understanding of the values predicted by the algorithm, and record a more solid result, one needs (at the very least) to perform multiple runs and determine the statistical variation of the obtained results. We expect the smallest statistical variations for the low-lying scaling dimensions and the corresponding OPE-squared coefficients. It would also be interesting to explore further how these data are affected by the choice of the z-sampling and the precise form of the reward function. As a preliminary check, we examined a derivative expansion of the crossing equations around the fully symmetric point $u=v=1$ (see Ref.~\cite{Li:2017ukc}), using the quoted scaling dimensions in Tab.~\ref{table:Mom_results_3} as an input. Truncating to the appropriate order we solved the resulting linear system to obtain the corresponding OPE-squared coefficients. Interestingly, we observed numerical values comparable to the ones obtained in Tab.~\ref{table:Mom_results_3} with the use of the RL algorithm.  

As an illustration, we performed a preliminary analysis of the statistical errors with multiple runs for the $\Delta_{\rm max} =2$, $\Delta_p  = 0.1$ case by completing 12 runs with 20 $z$-points in about 2 hours each. The results, collected in Tab.~\ref{table:Mom_results_statistical}, provide a more complete picture of the final output of the computation. We note that the errors in Tab.~\ref{table:Mom_results_statistical} do not include systematic errors associated with the truncation or the choice of the $z$-points. 

\begin{table}[t!]
\centering
\begin{tabular}{ | c | c || c | c || c | c |}
 \hline
\multicolumn{6}{|c|}{$\dmax=2$} \\
\hline
 Channel 			& spin 	& analytic $\Delta$ 	& RL $\Delta$ 	& analytic $\C$  & RL $\C$ 	\\ [0.5ex]
 \hline\hline
$s$ 				& 0 		& 0.4 			& 0.389413 $\pm$ 0.00862039 		& 1			&	0.995461 $\pm$ 0.00335572		\\ [1ex]
 \hline \hline
$t$			 	& 0		 & 2 				& 1.96776 $\pm$ 0.116733 			& 0.01	& 0.0115139 $\pm$ 0.00359425  		\\
				& 1		 & 1 				& 	0.96145 $\pm$ 0.0408496		& -0.1		& -0.101801 $\pm$ 0.00435701 		\\
 				& 2		 & 2 				& 	2.06592 $\pm$ 0.17467		&0.005		&0.00497925 $\pm$ 0.00156548 		\\ \hline \hline
\multicolumn{6}{|c|}{$\A = 0.000298727 \pm 0.0000960205$} \\ \hline
\end{tabular}
\caption{Analytic and numerical solutions from 12 runs for the mean and standard deviations of the scaling dimensions and OPE-squared coefficients for $\Delta_p=0.1$, spin-partitions with $\dmax = 2$ and 20 $z$-points.  The numerical results were obtained in mode 1.}
\label{table:Mom_results_statistical}
\end{table}

\begin{figure}[t!]
    \centering
    \includegraphics[width=0.5\textwidth]{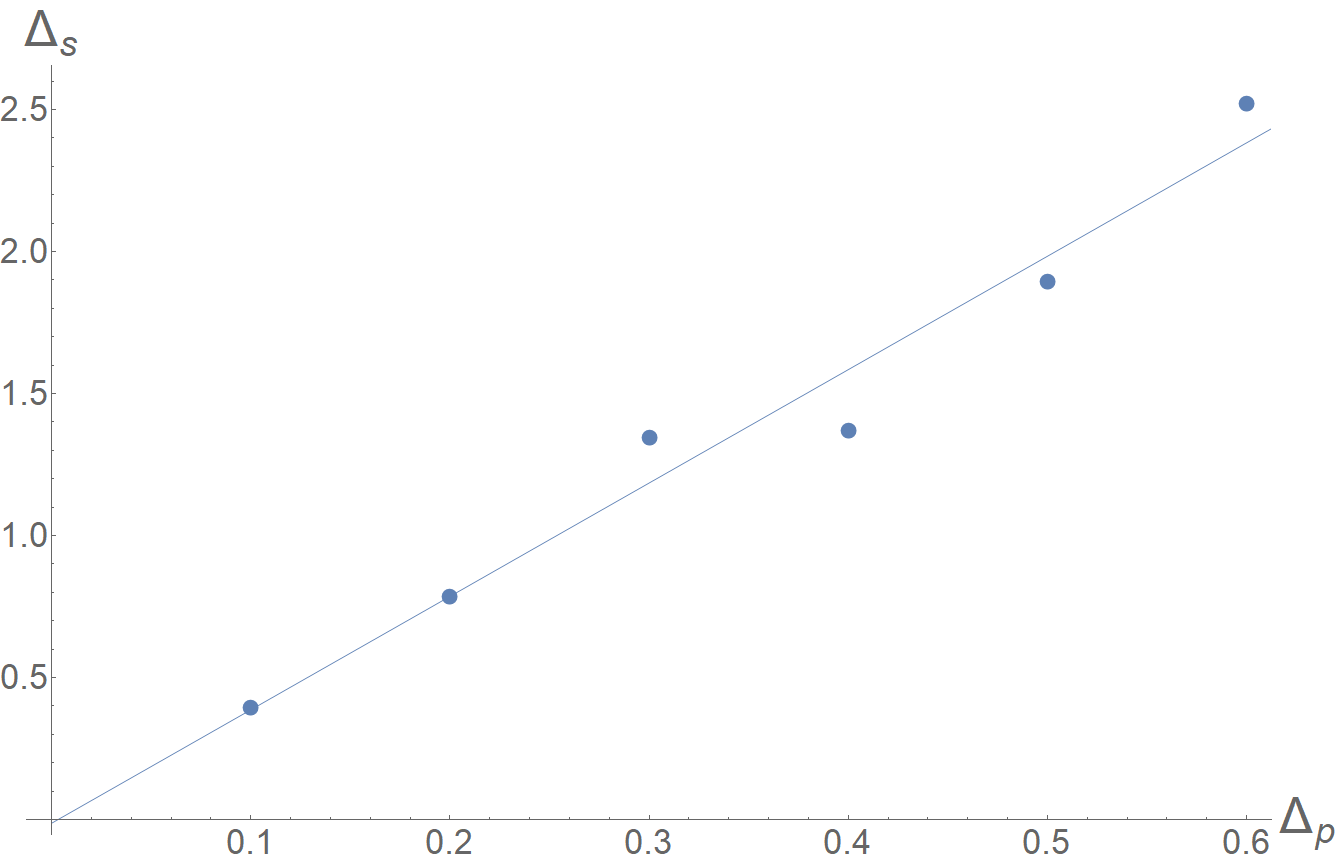}
    \caption{A plot of the numerically obtained values of $\Delta_s$ for the lowest scalar in the $s$ channel as a function of $\Delta_p$. The solid line is the line of best fit.}
    \label{ds_vs_dp}
\end{figure}

\begin{figure}[t!]
    \centering
    \includegraphics[width=0.5\textwidth]{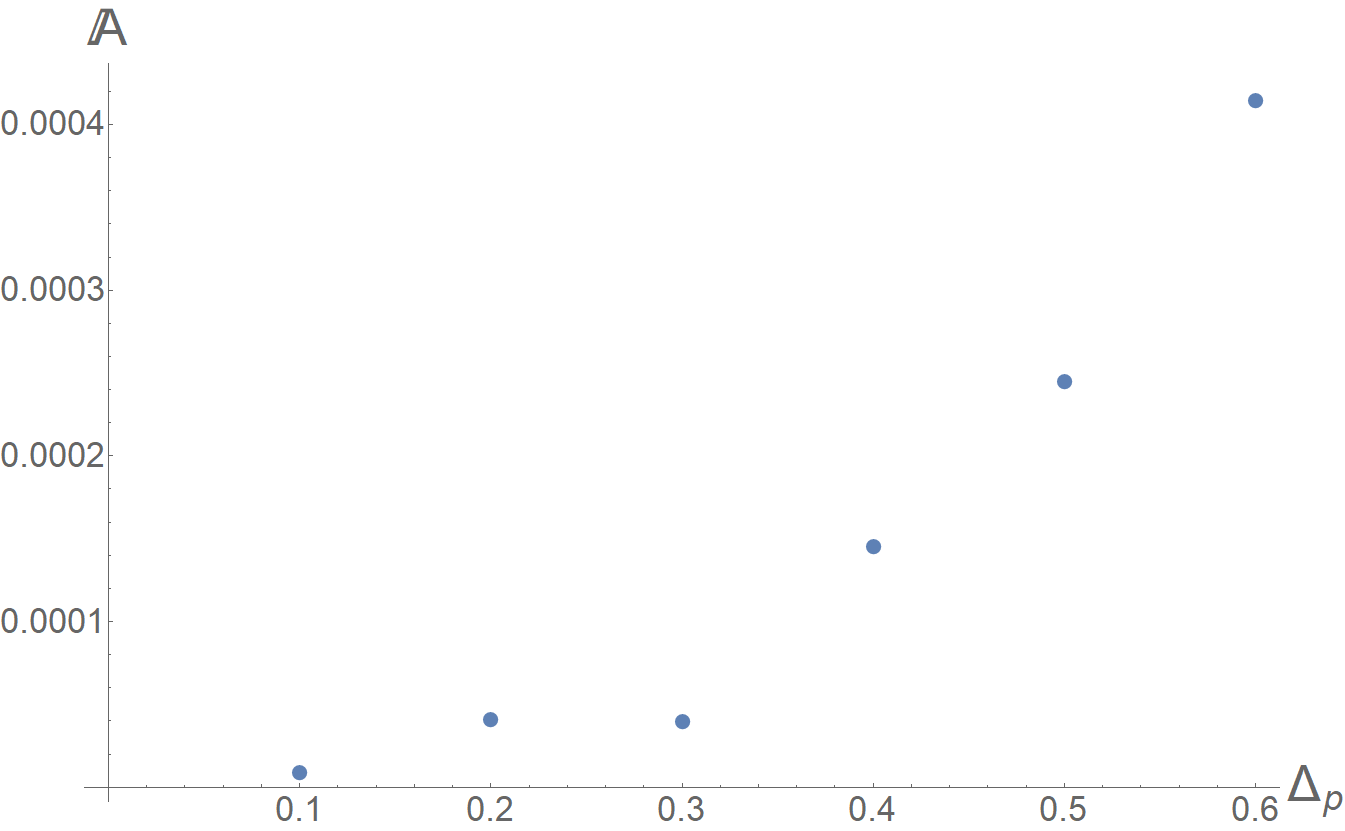}
    \caption{A plot of the relative accuracy $\mathbb A$ for the runs leading to Fig.~\ref{ds_vs_dp} as a function of $\Delta_p$.}
    \label{a_vs_dp}
\end{figure}

Finally, we performed the following exercise. Using the fixed spin-partition for $\dmax=2$ from Tab.~\ref{table:Mom_spin_partitions}, we varied $\Delta_p$ from 0.1 to 0.6 with a step of 0.1. As $\Delta_p$ increases so do the scaling dimensions in the $s$-channel. As a result, in the $s$-channel we increase appropriately the upper cutoff in the search and the fixed spin-partition is no longer that of $\dmax=2$. At the same time, the $t$-channel scaling dimensions remain within the $\dmax=2$ window. In Fig.~\ref{ds_vs_dp} we plot the scaling dimension $\Delta_s$ of the lowest scalar in the $s$-channel OPE of $V_p$ as a function of $\Delta_p$. The slope of the best-fit line, $\Delta_s = -0.0127 + 3.99345 \Delta_p$, is 0.16\% close to the analytically expected value of $\Delta_s = 4 \Delta_p$, although the relative accuracy $\A$ of the corresponding search increases for higher $\Delta_p$, as can be seen from Fig.~\ref{a_vs_dp}. This result suggests that the fixed spin-partition in the top entry of Tab.~\ref{table:Mom_spin_partitions} is inadequate as we increase the scaling dimension of the external operators and that more operators need to be included for large external scaling dimensions in both channels. It would be useful to develop a better understanding of the optimal use of cutoffs and search windows in such situations.  

One can also infer some additional information from Fig.~\ref{ds_vs_dp}. Had one been agnostic about the CFT, Fig.~\ref{ds_vs_dp} would provide evidence that the variable $p$ is proportional to the $U(1)$ charge of the operator $V_p$, since the scalar appearing in the OPE $V_p V_p$ has twice the $U(1)$ charge of $V_p$ (the $U(1)$ charge is additive) and the scaling dimension $\Delta_s$ is found to be $\Delta_s = (2p)^2$. A sharper argument along these lines could be obtained by studying the four-point function $\langle V_{p_1} V_{p_2} \bar V_{p_1} \bar V_{p_2}\rangle$ for a generic pair of $p_1, p_2$. The four-point function $\langle j j V_p \bar V_p \rangle$ would also yield related information.

At this point, it is interesting to ask whether the RL results allow us to conclusively determine that the CFT in question has a one-dimensional conformal manifold (namely an exactly marginal operator). The uncharged, spinless operator of scaling dimension 2 that appears in the $t$-channel is an obvious candidate that indicates the existence of a one-dimensional conformal manifold. Moreover, if there is some additional information that the spectrum of the CFT is discrete, the fact that we can solve the crossing equations for a continuous set of scaling dimensions $\Delta_p$ for the operators $V_p$, signals the fact that the theory has an exactly marginal deformation and that the scaling dimension of $V_p$ can be used as a proxy for the value of the exactly marginal coupling.

\begin{table}[t]
\centering
\begin{tabular}{ | c || c | c | c | c | c | c | c | c | c |}
 \hline
	Spin 		& 0 	& 1 	& 2 	& 3 	& 4	& 5	& 6	& 7	& 8	 \\ [0.5ex]
 \hline\hline
			& - 	& - 	& 1 	& -	& 1	& -	& 1 	& -	& 1	 \\ [1ex]
 \hline
\end{tabular}
\caption{A spin-partition inspired by the conformal-block decomposition of the four-point function $\langle j(z_1) j(z_2) j(z_3) j(z_4) \rangle$ with $\dmax=8$.}
\label{table:j_spin_partitions}
\end{table}

\begin{table}[t]
\centering
\begin{tabular}{ | c || c | c || c | c | c |}
 \hline
\multicolumn{5}{|c|}{$\Delta_j= 0.993577 \pm 0.00528402$} \\
 \hline
 spin 	& analytic $\Delta$ 	& RL $\Delta$ 	& analytic $\C$ & RL $\C$ 	\\ [0.5ex]
 \hline\hline
  2		& 2 				& $2.01688 \pm 0.0242115$	& 2			& $1.94164 \pm 0.0426005$	\\
  4		& 4 & $3.96686 \pm 0.0419861$ 		& 1.2			& $1.15899 \pm 0.0351261$	\\
  6		& 6 				& $5.95325 \pm 0.0438046$	& 0.23809524	& $0.22054 \pm 0.00729978$	\\
  8		& 8 				& $7.97585 \pm 0.0767531$	& 0.03263403	& $0.0240609 \pm 0.00121834$	\\ \hline \hline
\multicolumn{5}{|c|}{$\A = 0.000213657 \pm 0.00000819217$} \\ \hline
\end{tabular}
\caption{Analytic and numerical solutions from 10 runs for the mean and standard deviation of scaling dimensions and OPE-squared coefficients in the conformal-block decomposition of the four-point function $\langle j(z_1) j(z_2) j(z_3) j(z_4) \rangle$. $\Delta_j$ is also an unknown and the spin-partition is that of Tab.~\ref{table:j_spin_partitions}. The numerical results were obtained with 16 $z$-points and a mode-1 run of the RL algorithm.}
\label{table:j_stat_results}
\end{table}

\subsubsection{Spin-1 correlation functions}
\label{spin1}

A characteristic feature of the $S^1$ theory is the existence of a conserved holomorphic (and separately an anti-holomorphic) $U(1)$ current $j(z)$, under which many of the operators of the theory are charged. In this subsection, we study the four-point function of this current, \eqref{s1ac}. The holomorphic current $j(z)$ has spin 1 and (since it is conserved) scaling dimension $\Delta_j=1$. Keeping its scaling dimension $\Delta_j$ free for the moment, we find that the four-point function \eqref{s1ac} yields the crossing equation
\beq
\label{s1cb}
{\sum_{h \geq \bar h}}' \C_{h,\bar h} \Big( (z-1)^{2\Delta_j} g_{h,\bar h}^{(jjjj)}(z,\bar z) - z^{2\Delta_j}  g_{h,\bar h}^{(jjjj)}(1-z,1-\bar z) \Big) + \frac{1}{16} \Big( (z-1)^2 - z^2 \Big)  =0
~.
\eeq
The 1/16 factor in the last term, capturing the contribution of the identity, originates from the normalisation condition $\langle j j \rangle = \frac{1}{4}$.

The quasi-primaries that one needs in \eqref{s1cb} come from the  $j(z_1) j(z_2)$ OPE of the $S^1$ theory and can be straightforwardly obtained following the discussion around \eqref{s1analae}, by isolating contributions of the type $\partial^m X \partial^n X$  and setting $p\to 0$. These read
\begin{align}
  \label{eq:2}
 &(\d X)^2 ~, \qquad \d X \d^3 X - \frac{3}{2} (\d^2 X)^2 ~,\qquad \d X \d^5 X - 10 \d^2 X \d^4 X + 10 (\d^3 X)^2 \;,\cr
&\frac{1}{21}\partial X\partial ^7X -\partial ^2X\partial ^6X+ 5\partial ^3X\partial ^5X-\frac{25}{6}\left(\partial ^4X\right)^2\;,
\end{align}
and lead to the spin-partition of Tab.~\ref{table:j_spin_partitions} with $\dmax=8$.

With this spin-partition we ran the RL algorithm 10 times in mode 1 using 16 $z$-points. Each run lasted approximately two hours. In this case, we kept the conformal scaling dimension of the external operator $j$ as one of the unknowns to be determined by the agent. Overall, this was a 9-dimensional search. The results, collected in Tab.~\ref{table:j_stat_results}, include statistical errors and exhibit the relative accuracy $\A = (2.13657 \pm 0.0819217)\times 10^{-4} $. It is very rewarding to see that the agent determined the scaling dimension of the conserved $U(1)$ current to excellent accuracy just from the knowledge of the spin partition, and reproduced sensibly the low-lying spectrum and OPE data of the quasi-primary operators that appear in the OPE of the current with itself. For comparison, we also performed a single, independent mode-2 run with 16 $z$-points, where the scaling dimension of the current was fixed from the beginning at the analytic value $\Delta_j=1$. The results, at relative accuracy $\A = 0.00018822$, are summarised in Tab.~\ref{table:j_results}. They are nicely consistent with the mode-1 results of Tab.~\ref{table:j_stat_results}.

\begin{table}[t]
\centering
\begin{tabular}{ | c || c | c || c | c | c |}
 \hline
\multicolumn{5}{|c|}{$\Delta_j=1$} \\
 \hline
 spin 	& analytic $\Delta$ 	& RL $\Delta$ 	& analytic $\C$ & RL $\C$ 	\\ [0.5ex]
 \hline\hline
  2		& 2 				& 2.0080483	& 2			& 1.9766322	\\
  4		& 4 				& 4.0273294	& 1.2			& 1.1675011	\\
  6		& 6 				& 6.014957	& 0.23809524	& 0.21928822	\\
  8		& 8 				& 8.0047245	& 0.03263403	& 0.023595015	\\ \hline \hline
\multicolumn{5}{|c|}{$\A = 0.00018822$} \\ \hline
\end{tabular}
\caption{Analytic and numerical solutions for scaling dimensions and OPE-squared coefficients in the conformal-block decomposition of the four-point function $\langle j(z_1) j(z_2) j(z_3) j(z_4) \rangle$ for $\Delta_j=1$ and the spin-partition of Tab.~\ref{table:j_spin_partitions}. The numerical results were obtained with 16 $z$-points and a mode-2 run of the RL algorithm.}
\label{table:j_results}
\end{table}

\section{Conclusions and Outlook}
\label{outlook}

In this paper we introduced the use of Reinforcement-Learning techniques into the conformal bootstrap programme. We tested an RL soft Actor-Critic algorithm in the context of several 2D CFTs and showed that the algorithm can perform efficient multi-dimensional searches in the space of scaling dimensions and OPE-squared coefficients. The basic input of our approach is a spin-partition and a window of scaling dimensions, where the search is concentrated. We demonstrated in concrete examples that this minimal input is enough to guide the algorithm towards a CFT of interest and that the obtained numerical values can be sensible even in very rough truncations with only a handful of operators. Our algorithm can be straightforwardly applied to any CFT of arbitrary spacetime dimension. This opens up the very exciting possibility of new non-perturbative results in conformal field theory in a wide range of directions, some of which we plan to explore in the near future.

We view the approach introduced here as largely complementary to the more standard ones that have already been developed to-date in the context of the numerical conformal bootstrap. We believe that our method is comparatively stronger in performing efficient multi-dimensional searches in arbitrary, a priori selected (unitary or non-unitary) CFTs. Since it is based on statistical and probabilistic techniques, it can be weaker in accuracy, on detecting rigorous bounds and on conclusively rejecting CFT data as inconsistent. The latter is the context where standard numerical conformal-bootstrap approaches have excelled over the last decade. Eventually, one would like to combine all available analytic and numerical methods at their disposal to build a powerful multi-purpose toolbox.

We envisage the most efficient application of our approach in contexts where a CFT can be solved in a parametrically convenient regime (e.g.\ in a weakly coupled large-$N$ regime or a weakly coupled regime on a conformal manifold). Then, one can use the information of the perturbative solution to set up a well-informed spin-partition, that can in turn be applied adiabatically to a search with gradually changing parameters. By using a gradual update of the CFT data, one should be able to implement the RL algorithm step-by-step and track them from a weak- to a strong-coupling regime. This is a concrete context, where one can try to leverage all available analytic and numerical information. For example, in superconformal field theories, our approach can benefit from many recent developments that use the superconformal structure of the theory in an essential way.

Although our results provide a proof of principle for the usefulness of RL techniques to this class of problems, there are several aspects of our approach that require further investigation and development. The most urgent is to systematically understand  how to incorporate reliable errors in our computations. The primary source of error is of an analytic nature and originates from the truncation of the conformal-block expansions. The convergence properties of these expansions, \cite{Pappadopulo:2012jk}, imply that there is a sufficiently high $\dmax$ above which the error will be negligible. It is unclear, however, how to identify this optimal $\dmax$ in a generic theory and for generic four-point functions. Hence, one might initially need to perform a case-by-case analysis in order to explore how our results are affected by an increasing $\dmax$.

Another source of error, which is sometimes more significant than the error due to the $\dmax$ truncation, comes from the way we reduce the functional dependence of the crossing equations on the cross-ratios to a discrete set of algebraic equations. In this paper we have chosen to implement this reduction by evaluating the crossing equations on a finite set of cross-ratio values. We noticed experimentally that the sampling of $z$-points suggested in Sec.~3.1 of \cite{CastedoEcheverri:2016fxt} works well in our computations. However, we lack a good understanding of whether this is the optimal sampling, or how the calculations are affected by the number of $z$-points selected. An error can consequently be associated with these effects by varying the sampling (in form and size). Alternatively, one can explore more standard reductions based on Taylor-expansions of the conformal blocks around some point in $z$-space. It would be interesting to repeat the computations of this paper with this alternative approach and compare results.

Other errors have to do with the statistical nature of our approach and the fact that we do not a priori know the minimal possible violation of the truncated crossing equations for a given truncation and reduction. In this paper we quantified this violation with a relative measure of accuracy $\A$ and performed runs of the RL algorithm up to the point where the improvement of $\A$ was saturated. An important additional measure of error for each CFT datum is a statistical error obtained by performing the same type of run many times, which we sampled in the case of the $c=1$ compactified boson CFT on $S^1$ for the simplest case of $\dmax = 2$ in the momentum sector and for $\dmax=8$ in the four-point function of the conserved $U(1)$ current. The evaluation of this type of error would benefit from a fully parallelisable algorithm. As we noted in Sec.~\ref{modes}, current implementations of the algorithm benefit from the judicious caretaking of the user, which obstructs the full parallelisability of the code. It would be useful to improve this aspect in future work.

In this paper we did not make systematic use of the constraints of global symmetries or of the full constraints of unitarity on the OPE-squared coefficients. As we observed in Sec.~\ref{momentumsector}, multi-dimensional searches can benefit significantly from prior information on the signs of the OPE-squared coefficients. Without such information the agent is allowed to explore cancellations between different conformal blocks that sidetrack the search by increasing the statistical error on certain OPE-squared coefficients, especially so for those at higher scaling dimensions that come naturally with suppressed numerical values.

Finally, we treated the learning algorithm itself as a black box, using the off-the shelf soft Actor-Critic algorithm of \cite{DBLP:journals/corr/abs-1801-01290}. It would be interesting to explore what efficiency and speed gains one can achieve by tuning hyperparameters or choosing the Deep Deterministic Policy Gradient method \cite{Lillicrap2016ContinuousCW}. We also chose the simplest definition for the reward function \eqref{eq:1}. The choice of an appropriate reward function is crucial in achieving better results for RL algorithms and this is an area that also deserves further investigation.

%%%%%%%%%%%%%%%%

\ack{ \bigskip We would like to thank P.~Agarwal, E.V.G.~Christiansen, S.~Kousvos, B.~van Rees and P.~Richmond for useful discussions and comments. C.P.\ is supported by the Royal Society and the STFC under grant numbers  UF120032 and ST/P000754/1.}

%%%%%%%%%%%%%%%%

%% Bibliography

\bibliography{machine_learning}

\end{document}